\documentclass[12pt,preprint]{aastex}
\usepackage{graphicx,epsfig}
\usepackage{lscape}
\usepackage{rotating}
\newcommand{\bq}{\begin{equation}}
\newcommand{\eq}{\end{equation}}

\def\gtsim{\lower.5ex\hbox{$\buildrel > \over\sim$}}
\def\ltsim{\lower.5ex\hbox{$\buildrel < \over\sim$}}

\def\apjl{ApJL}
\def\apj{ApJ}
\def\apjs{ApJS}
\def\mnras{MNRAS}

\def\aj{AJ}
\def\aap{A\&A}

\def\nat{Nature}

\shorttitle{SN Light Curves}
\shortauthors{Chatzopoulos,Wheeler,Vinko}
\begin{document}
\title
{Generalized Semi -- Analytical Models of Supernova Light Curves}
\author{E. Chatzopoulos\altaffilmark{1}, J. Craig Wheeler\altaffilmark{1} \& J. Vinko\altaffilmark{2,1}}
%%%  author names
\authoremail{manolis@astro.as.utexas.edu}
\altaffiltext{1}{Department of Astronomy, University of Texas at Austin, Austin, TX, USA.}
\altaffiltext{2}{Department of Optics and Quantum Electronics, University of Szeged, Hungary}

\begin{abstract}
 
We present generalized supernova (SN) light curve (LC) models for a variety of power inputs including the previously proposed ideas
of radioactive decay of $^{56}$Ni and $^{56}$Co and magnetar spin-down. We extend those solutions to include finite progenitor radius
and stationary photospheres as might be the case for SN that are powered by interaction of the ejecta with circumstellar 
matter (CSM).
We provide an expression for the power input that is produced by self-similar forward and reverse shocks that efficiently convert 
their kinetic energy into radiation. 
We find that this ejecta-CSM interaction luminosity that we derive is in agreement with results from multi-dimensional radiation 
hydrodynamics simulations in the case of an optically-thin CSM. We develop a semi-analytical model for the case of an optically-thick 
CSM by invoking an approximation for the effects of radiative diffusion similar to that adopted by Arnett (1982) for SN II 
and compare this model to the results of numerical radiation hydrodynamics models.
This model can give complex light curves, but for monotonically declining shock input, the LCs have a smooth rise, peak and
decline.
In the context of this model, we provide predictions of the shock breakout of the forward
shock from the optically-thick part of the CSM envelope. We also introduce
a hybrid LC model that incorporates ejecta-CSM interaction plus $^{56}$Ni and $^{56}$Co radioactive decay input. We fit this hybrid model 
to the LC of the Super-Luminous Supernova (SLSN) 2006gy. 
We find that shock heating produced by ejecta-CSM interaction plus some contribution from radioactive decay 
provides a better fit to the LC of this event than previously presented models. 
We also address the relation between SN IIL and SN IIn with ejecta-CSM interaction models. 
The faster decline of SN IIL can be reproduced by the diffusion of previously 
deposited shock power if the shock power input to the diffusive component vanishes when the reverse shock sweeps up the whole ejecta and/or
the forward shock propagates through the optically-thick CS matter.
A CSM interaction with forward and reverse shock power input can produce the LCs of SN IIn in terms of duration, shape and decline rate, 
depending on the properties of the CSM envelope and the progenitor star. This model can also produce LCs that are symmetric in shape around peak
luminosity, which is the case for the observed LCs of some recently discovered peculiar transient events.
We conclude that the observed LC variety of SN IIn and of some SLSNe is likely to be 
a byproduct of the large range of conditions relevant to significant ejecta-CSM interaction as a power source.
 
\end{abstract}

\keywords{circumstellar matter -- stars: light curve -- supernovae: general -- supernovae: individual -- light curve}

\vskip 0.57 in

\section{INTRODUCTION}\label{intro}

The current SN classification scheme is based on both the properties of the spectrum and of the light curve. 
The basic properties of stripped envelope SNe (Ia,b,c) light curves (LCs) are well reproduced by considering the diffusion 
of the radioactive decay energy of $^{56}$Ni and $^{56}$Co
into homologously expanding SN ejecta (Arnett 1979, 1980, 1982, 1996 - hereafter A79, A80, A82, A96). Core-collapse SNe are divided
into the Type Ib/c and Type II subclasses. SN Ib/Ic are considered explosions of compact progenitors that have
lost their outer hydrogen (for SN Ib) and helium (for SN Ic) envelopes. A79 addressed what we would now call SN Ib; A80 and A82
pertained more directly to SN Ia, but the principles are the same.
For Type II SN, the following subtypes have been proposed: Type IIP, Type IIb, Type IIL and Type IIn. 
SN IIP explosions are the most common, and they are believed to be the result of the death of a massive red supergiant (RSG) progenitor star.
The LCs of SN IIP are characterized by a long plateau that is indicative of the energy
liberated by a recombination front in the extended hydrogen envelope of the progenitor star. The late time decline rate
of many SN IIP LCs is consistent with that of the radiactive decay of $^{56}$Co. Arnett \& Fu (1989) presented
a semi-analytical model that incorporates the effects of radiative diffusion with energy deposition from the radioactive decay of $^{56}$Co 
(also pulsar and fallback energy) plus H recombination in order to fit the observed LC of SN~1987A. 
This model can produce a variety of SN IIP-like LCs (and probably
SN IIb, although this has not been explored in depth)
depending on the choice of H mass and the composition of the outer shell of the progenitor RSG star. More
accurate radiation hydrodynamics simulations of SN IIP have been done yielding simultaneous LCs and spectra
(see examples in Falk \& Arnett 1977; Falk 1978; Klein \& Chevalier 1978; Kasen \& Woosley 2009; Dessart, Livne \& Waldman 2010; 
Bersten et al. 2011; Dessart \& Hiller 2011) as well as models of 
SN~1987A that have reproduced the observables in great detail (Blinnikov et al. 2000; Dessart \& Hiller 2010).
One-dimensional numerical LC models based on radiation hydrodynamics calculations are available for SN IIL 
(Swartz et al. 1991; Blinnikov \& Bartunov 1993),
but a simple semi-analytical approach, similar to that presented by A82, A96 for the case of SN Ia has not been provided. The simulations by
Swartz et al. (1991) were able to reproduce some of the characteristics of SN IIL LCs under the assumption of O-Ne-Mg (Oxygen-Neon-Magnesium)
core-collapse, but the fits were not good for all phases of the observed LCs.
The LCs of SN IIn present a diversity when it comes to shape, peak luminosity, duration
and decline rate. Some of the recently discovered Super-Luminous Supernovae (SLSNe) are classified as SN IIn events, 
mainly due to the spectroscopic signatures of ejecta-CSM
interaction (Schlegel 1996). 

Although there are numerical simulations of the interaction of SN ejecta with the CSM and model LCs under optically-thin
conditions (Falk \& Arnett 1977; Van Marle et al. 2010) the only attempt made so far to incorporate the effects of radiative diffusion of 
shock-deposited energy in an optically-thick CSM envelope 
was done in the case of red supergiant (RSG) progenitors surrounded by a CSM with
a limited set of density profiles ($\rho_{CSM} \sim r^{-2,-1.5}$) (Moriya et al. 2011). The work done by Moriya et al. (2011)
using the 1-D radiation hydrodynamics code STELLA (Blinnikov \& Bartunov 1993; Blinnikov et al. 1998; Blinnikov et al. 2006)
reproduced the LC of ultraviolet-rich SN IIP SN~2009kf reasonably well and provided a tool to study the LCs of low-luminosity SN IIn
powered by the interaction of the SN ejecta with a CS wind. For higher luminosity events in the regime of SLSNe, 
Moriya et al. (2011) find that extraordinary mass-loss rates are required that are inconsistent with an RSG progenitor. 
This and other factors motivate us to update the analytical models of A80, 82 and determine approximate LCs for SLSNe 
powered by optically-thick SN ejecta - CSM interaction for a variety of CSM and progenitor characteristics. 
This is important in order to produce a physically acceptable model for a SLSN IIn LC for which there is a smooth rise
to maximum light; a simplified model of instantaneous shock heating yields a LC that increases steeply to maximum in a very short time-scale.
The large parameter space that is connected with SN IIn 
(properties of the SN progenitor, the SN ejecta and of the CSM envelope) 
provide a natural explanation for the observed diversity of SN IIn LCs. Another motivation for developing an efficient approximate analytical
model is to perform a qualitative study of this large parameter space.
Given that many of the recently discovered SLSNe have been classified as
SN IIL or SN IIn events, a ``unified" model is sought in order to explain their properties; 
at least the properties of their CSM environments and ideally
of their progenitors as well.

One of the interesting aspects of the SLSN events that seem to be surrounded by dense, optically-thick shells is that the
nature of the underlying supernova is hidden and thus remains obscure. There seems to be too much mass to be associated
with any underlying white dwarf thermonuclear explosion, but that leaves many unanswered questions. What are the progenitor
mass and radius, the ejecta composition and energy? What is the explosion mechanism, core-collapse or something else?
The collective enigmatic nature of these super-luminous events means that we have to return to rudimentary studies
to explore the parameter space that may be appropriate. For this reason we have generalized the basic light curve models of
A80, 82 using the first law of thermodynamics and the diffusion approximation. Where Arnett first proposed his
models in the context of radioactive decay, we have adapted his technique for an arbitrary prescription of the power input.
This allows us to explore radioactive decay, shell-shocks, magnetars and, within some limitations, combinations of these
power inputs. These basic models allow us to address questions such as: what is needed to make the light curve rise have a
certain shape, and, independently, what shapes the decline; what is needed to generate a nearly symmetric light curve, or a
very asymmetric one; what is the effect of the initial radius of the supernova progenitor;
what are the constraints on the density profiles in the supernova ejecta and the circumstellar matter? The most general question
we propose to explore is whether or not the heterogeneity of the current sample of SLSN requires very different progenitors and physics,
or if there is some common theme expressed in different ways. The goal of this study is not to compete with more 
realistic radiation hydrodynamics simulations but rather complement them by providing approximate solutions that will help us understand
the importance of the parameters involved in the case of SLSNe. Benchmarking against more accurate, numerical results is
discussed in this work as a means to illustrate the uncertainties and the limitations of our model.

We organize the paper as follows: In \S2 we present the general physical assumptions of the basic LC model
and we provide solutions in the cases of homologously expanding material and fixed SN photosphere. 
In \S3 we present a variety of physically-motivated luminosity inputs 
and we develop model LCs for shock energy deposition resulting from SN ejecta-CSM interaction incorporating the effects of radiative diffusion. 
We also develop a hybrid LC model with ejecta-CSM interaction plus $^{56}$Ni and $^{56}$Co radioactive decay.
In \S4 we provide a characteristic fit of the hybrid ejecta-CSM interaction radioactive decay input
to the LC of the SLSN~2006gy and discuss the implications of this model for the nature of the event. Applications to other SLSNe
will be presented in a subsequent paper.
Finally, in \S5 we summarize our conclusions. Details of the derivations of the analytical models are given in Appendices A and B. 

\section{GENERAL LIGHT CURVE MODEL}\label{lcgen}

 Following the prescriptions of A80, 82 for Type I SN LCs,
we present a solution for a general heating input. The initial assumptions for our model remain the same 
as in A80, 82, those being:
1) homologous expansion of the ejecta
2) centrally located power input source
3) radiation pressure dominant.
Under these assumptions we consider the first law of thermodynamics:
\begin{equation}
\dot{E}+P \dot{V}=\epsilon_{inp}-\frac{\partial L}{\partial m},
\end{equation}
where $E=aVT^{4}$ is the specific internal energy, $P=(1/3)aT^{4}V$ is the pressure, $V=\rho^{-1}$ is the specific
volume where $\rho$ is density, $\epsilon_{inp}$ is the specific input energy generation rate,
$L$ the output radiated luminosity and $m$ is the mass coordinate of the fluid element. 
In general, the temperature profile of the diffusion mass, $T$, $\epsilon_{inp}$ and $L$
are functions of position, $x=r/R$, and time, $t$, where $x$ is the dimensionless 
position coordinate for a fluid element relative to a fiducial radius taken to be the radius of the photosphere. For homologous
expansion $R=R_{0}+v t$ where $R_{0}$ is the initial radius of the ejecta at the moment of shock breakout 
and $v$ is the characteristic expansion velocity of the ejecta. The velocity $v$
is not necessarily the photospheric expansion velocity, $v_{ph}$, as measured from SN spectra,
but we can use $v=v_{ph}$ as an approximation in some cases. 

For the output luminosity we use the radiation diffusion approximation
\begin{equation}
L=-\frac{4 \pi r^{2} \lambda c a}{3} \frac{\partial T^{4}(x,t)}{\partial r},
\end{equation}
where $\lambda=1/\kappa \rho$ is the mean free path with $\kappa$ being the mean
opacity that we take to be a constant, $\rho$ the density, and $c$ the speed of light.
In the following analysis we alter the two first criteria of A80, 82 by considering $v \rightarrow$~0 (fixed photospheric radius instead
of homologous expansion) and input sources that terminate due to their movement
through the diffusion mass such as a forward shock that breaks out from an optically-thick CSM envelope.
The assumption of constant opacity is a weakness of our analytical approach, but it was necessary in order to obtain
a separable PDE. In more realistic situations that are accounted for in radiation hydrodynamics models opacity is depth and time dependent. 
We note that while Moriya et al. (2011) use their radiation hydrodynamics code to compute the ionization state and opacities
of the underlying explosion, they adopt a Thomson electron scattering opacity for fully ionized solar metallicity material
($\kappa \sim 0.33$~cm$^{2}$~g$^{-1}$) within the ionization front of the CSM.
We will use the same value for the optical opacity throughout this work.
This general LC model has been considered and expanded by Blinnikov \& Popov (1993) for the case
of a piecewise constant opacity in power-law density distributions for the SN ejecta. As pointed out in their paper,
the correct approach to the subject belongs to the class of moving boundary problems and not the usual eigenvalue formulation that has been
considered so far for the analytical models. A different approach has also been considered by Popov (1995) where he adopts
a mixed boundary condition for radiation (not the radiative zero solution that is considered in the A80, A82, A96 models) and obtains
solutions for general heating terms in the form of Green's functions but specifically focusing attention on radioactive heating.

\subsection{{\it Solution for homologously expanding photosphere}}\label{homolexp}

As shown in Appendix A, the general full solution for the output luminosity from the photosphere of the SN ejecta
can be written as:
\begin{equation}
L(t)=\frac{2L_{0}}{t_{d}} e^{-[\frac{t^{2}}{t_{d}^{2}}+\frac{2R_{0}t}{vt_{d}^{2}}]}
\int_0^t e^{[\frac{t'^{2}}{t_{d}^{2}}+\frac{2R_{0}t'}{vt_{d}^{2}}]} f(t')
[\frac{R_{0}}{vt_{d}}+\frac{t'}{t_{d}}]dt' + \frac{E_{th,0}}{t_{0}} e^{-(t^{2}/t_{d}^{2}+2R_{0}t/vt_{d}^{2})},
\end{equation}
where we have introduced the effective light curve time scale, $t_{d}=\sqrt{2t_{0}t_{h}}$, and 
where $t_{0} = 3 \kappa R_{0}^{2}/V_{00} \alpha c = \kappa M /\beta c R_{0}$ is the diffusion time scale, $V_{00}$ is the initial central specific
volume of the ejecta, $\alpha$ is a constant arising from the separation of variables of the LC PDE,
$t_{h}=R_{0}/v$ is the expansion time-scale, $L_{0}$ is the initial luminosity input (see Appendix A),
$E_{th,0}$ the total SN explosion energy and $\beta$ is a constant that accounts
for the density profile of the diffusion mass. A80, 82 adopt $\beta =$~13.8 as a good approximation for a variety of diffusion mass density profiles.
The term $R(t)/R_{0}t_{0}$ (see Equation A9 in Appendix A) can be written as $(R_{0}+vt)/R_{0}t_{0}$ or 
$(2R_{0}/vt_{d}+2t/t_{d})/t_{d}$ (as we have done in Equation 3) as convenient. 
For $R_{0} \rightarrow 0$ the additive term $R_{0}t/vt_{d}^{2}$ that appears in Equation 3 vanishes and the solution for small initial radius is recovered, as presented
in A80, 82. Arnett \& Fu (1989) and A96 also implicitly present this general solution and apply it to the case of the LC of SN~1987A
for several power inputs (pulsar heating, swept-up luminosity and fallback).
The second term in Equation 3 is an initial value term related to the internal energy, $E_{th,0}$ that the SN 
possesses at a given time. In Equation 3 this term is governed by the energy the ejecta have at time $t =$~0 
that they gained from the SN blast wave that subsequently
diffuses from the optically-thick expanding envelope. In A80, this term is used to explain SN IIP LCs, for which the CSM shock input
energy from radioactive $^{56}$Ni and $^{56}$Co decay is small compared to the total energy of the ejecta at early times. 
Due to the fast exponential decay this ``fireball" term may be ignored for our purposes 
since it affects the LC only at very early times. We note that
this term does play a role in models for which the input ceases and diffusion controls the luminous output.
In this context, the coefficient of the exponential term becomes $L_{0}$, the initial value of the luminosity at the time the power input ceases, after
which the luminosity decays according to the exponential diffusion term (\S3.1, \S3.2.1).
In \S3 we will explore several physical luminosity inputs to obtain the final generalized model LCs.

\subsection{{\it Solution for fixed photosphere}}\label{fixedph}

Next, rather than homologous expansion, we assume $v=0$, $R=R_{0}=R_{ph}$ (and thus $R(t)/R_{0}t_{0}=1/t_{0}$) for
a fixed photosphere radius. The motivation for this is the fact that around massive SN progenitors 
there can be an optically-thick CSM shell, so that when 
the SN ejecta collide with that surrounding medium, a forward shock is formed that propagates
into the nearly stationary CSM and a reverse shock propagates into the ejecta, both depositing kinetic energy and
heating the interacting media. These shocks provide a natural source for the output luminosity of the event 
(Chevalier 1982; Chevalier \& Fransson 1994; Chugai \& Danziger 1994). Although the diffusion problem that we solved above involves an homologously 
expanding SN photosphere, we assume the same principles hold for a stationary photosphere and diffusion mass.
This assumption modifies the final PDE that we solve (see Appendix A) and leads to the following general solution:
\begin{equation}
L(t)=\frac{1}{t_{0}} e^{-\frac{t}{t_{0}}} \int_0^t e^{\frac{t'}{t_{0}}} L_{inp}(t')dt'+\frac{E_{th,0}}{t_{0}}e^{-t/t_{0}},
\end{equation}
where $E_{th,0}$ is, again, the initial internal energy from the SN blast wave that affects the LC at early
times and that we ignore in our analysis. For small initial radius ($R_{0} \rightarrow 0$) that term also goes to zero.

\section{POWER INPUTS}\label{inps}

Having the general solutions for the output SN LCs provided by Equations 3, and 4 for homologously expanding and
fixed photosphere, respectively, we now consider several physical heat inputs 
as the ``source functions" for the integrals of these expressions in order to obtain the final output LC models. 
From the form of these general solutions it follows that, in order to obtain a physical LC that rises smoothly to maximum 
and subsequently declines, the luminosity deposition function must be a smooth continuous function that does not monotonically increase.
For a constant or monotonically rising power input, the output luminosity will always increase with time with a monotonically increasing slope
until the heat input vanishes at some point. For a monotonically declining input,
the luminosity will increase to a maximum value and then decline, dominated by diffusion and cooling.
As shown in \S3.3, plausible input sources can rise and then decline.
At very late times, the output luminosity will be the same as the input luminosity as the diffusion time becomes
short compared to the total elapsed time. This property is used as a diagnostic for radioactive decay of $^{56}$Ni and $^{56}$Co for some SN LCs
where late-time photometric observations indicate a decline rate consistent with that of $^{56}$Co 
(Colgate \& McKee 1969; Colgate et al. 1980). The same is expected to be the case for other continuous power inputs.

We may consider two general categories of power inputs: centrally located ones and moving ones.
A well-known input is the radioactive decay of $^{56}$Ni and $^{56}$Co that is believed to
power the SN Ia, SN Ib/c and some SN II LCs (A80, A82), particularly the maximum of SN~1987A and SN IIb. 
Although there can be some outward mixing,
the newly-formed nickel is often taken to be confined near the center
of the SN ejecta, around the core of the progenitor star. This assumption agrees well
with some simulations of SN Ia, SN Ib/c and SN II events (Nomoto et al. 1984). Another centrally located input that has been 
considered for SNe is radiation from a magnetic dipole associated with a pulsar (Ostriker \& Gunn 1971) as applied
in the case of SN~1987 by Arnett \& Fu (1989) or a magnetar as
proposed recently as the power source for the SLSNe 2007bi and 2008es (Kasen \& Bildsten 2010;
Woosley 2010; see also Maeda et al. 2007).

Moving inputs include the energy released in SN ejecta due to the recombination of H and He
and shock heating. Arnett \& Fu (1989) presented a semi-analytical model that takes the effects of H
recombination into account in order to reproduce the LC of SN~1987A. The recombination front recedes 
into the expanding SN ejecta. The effect of that is the creation of a plateau phase in the SN LC that is
more pronounced for large initial radius and large mass of the H envelope. The analytical
model presented in Arnett \& Fu (1989) can therefore be used to reproduce some SN IIP LC characteristics. 
Analytical models for SN IIP LCs have also been provided by Popov (1993) and Kasen \& Woosley (2009).

Shock heating can be another important moving power source for some SN LCs. 
SN IIn show evidence for CSM interaction
in their spectra with prominent H and sometimes He emission features indicating the presence of a shock running
into the CSM and depositing kinetic energy (Chevalier 1982; Chevalier \& Fransson 1994). A significant fraction
of the recently discovered SLSNe are SN IIn and show signs of CSM interaction, so an analytical model
that describes the output light curve that results from this power input might be helpful in understanding the basic properties
of the CSM involved in the process. 
A peculiarity of this input is that its dynamics depend on the physical characteristics of the interacting
media (the SN ejecta and the CSM shell). In addition, it is not a continuous input; once the reverse shock sweeps up
all the available ejecta mass or the forward shock sweeps up all the available CSM mass
there is no further heating and all that is left is just the diffusion of the previously shock-deposited
energy. Incorporating forward shock heating into an optically-thick CSM envelope requires a different treatment
in order to account for the movement of the input source, and the resulting PDE is not formally separable. For the scope
of this work, we will assume that reverse and forward shock heating are both centrally located, but that they terminate; forward
shock heating terminates when the shock breaks out of the CSM and reverse shock heating when the available SN ejecta
mass has been swept-up. 
With these assumptions, the PDE becomes separable (Appendix B). 
The assumption of centrally located power source for the case of the forward and the reverse shocks, although convenient,
is not generally true
and thus increases the uncertainties and limitations of this approximate model. For this reason in \S3.3.4 we compare our results with numerical
results presented by Chugai et al. (2004), Woosley, Blinnikov \& Heger (2007) and Moriya et al. (2011) for the same initial conditions and discuss the 
differences. These approximate solutions give useful guidance, as we will see, but
a proper solution calls for numerical radiation hydrodynamics.
A combination of power sources, centrally located and moving/terminated,
is probably involved for most core-collapse SN. We consider such a case by developing a hybrid radioactive decay and
CSM interaction diffusion model in \S3.3.3.

\subsection{{\it Instantaneous luminosity input at the SN photosphere}}\label{inst}

We start by considering the simple initial value problem where the initial luminosity $L_{0}$  
is introduced at $t=t_{max}$ at the SN photosphere. This problem was also considered by A80 and deals
with the resulting SN LC if the energy input were just the initial SN shock energy for which $L_{0}=E_{0}/t_{0}$ in Equation 3. 
This input produces a light curve with instantaneous rise to maximum light at $t = t_{max}$ followed by 
simple diffusion. A96 also presented solutions of this ``expanding fireball" problem  
that has the following form in the homologous expansion case:
\begin{equation}
L(t)=L_{0} e^{-[\frac{(t-t_{max})^{2}}{t_{d}^{2}}+\frac{2R_{0}(t-t_{max})}{vt_{d}^{2}}]}.
\end{equation}
In the fixed photosphere case, this reduces to: 
\begin{equation}
L(t) = L_{0} e^{-(t-t_{max})/t_{0}}.
\end{equation}
These results are the same in form as the second terms of Equations 3 and 4 for homologous expansion and fixed photospheric radius,
respectively. Pure diffusion on the decline has the same form for any luminosity input that vanishes after some time interval.
Similar diffusive decay from an initial value of luminosity at a given time pertains to models in which the power input is truncated (\S3.2.1, \S3.3.2).
This simple solution has been considered by some authors as an interpretation for the decline of some SLSNe
(for an example, see Smith \& McCray 2007 for the LC of SN~2006gy). Model LCs powered by initial value luminosity
input are shown in Figure 1 for the cases of homologously expanding matter and stationary photospheres.
This model does not provide a natural explanation for a LC with a smooth extended rise.

\subsection{{\it Centrally located power inputs}}\label{ceninp}

Centrally located power inputs such as the radioactive decay of $^{56}$Ni and $^{56}$Co as well
as magnetar spin-down have been considered in previous work. In the following subsections we
summarize those past results in order to use them for comparison with the models that we will introduce.

\subsubsection{{\it Terminated constant CS shock luminosity}}\label{tophat}

 Motivated by the fact that an instantaneous shock luminosity input does not provide an output SN LC with a smooth rise to maximum light,
we consider a ``top-hat" model in which a constant input is provided for a finite time and then shuts off. To be specific, we consider the input to be a
shock with constant luminosity, $L_{sh}$ lasting for time, $t_{sh}$, and thus producing total energy $E_{sh}=L_{sh}t_{sh}$.
Therefore, the input shock luminosity that is provided to the expanding diffusion mass is $L_{sh} = E_{sh}/t_{sh}$ for $t<t_{sh}$
and $L_{sh} = 0$ otherwise. This is the input luminosity function that we insert into Equations 3 and 4 to calculate the output LC in the cases 
of homologously expanding matter and stationary matter with a photosphere of fixed radius, respectively. We also assume that the shock luminosity deposition
takes place deep within the diffusion mass, therefore we neglect any corrections attributable to the movement of the shock towards the photosphere. 
Direct integration yields
\begin{equation}
L(t) = \cases{\frac{E_{sh}}{t_{sh}} [1-e^{-(t^{2}/2t_{d}^2+2R_{0}t/v t_{d}^{2})}], & $t < t_{sh}$, \cr
\frac{E_{sh}}{t_{sh}} e^{-(t^{2}/2t_{d}^2+2R_{0}t/v t_{d}^{2})} [e^{(t_{sh}^{2}/2t_{d}^2+R_{0}t_{sh}/v t_{d}^{2})}-1], & $t > t_{sh}$,}
\end{equation}
in the case of homologously expanding matter and
\begin{equation}
L(t) = \cases{\frac{E_{sh}}{t_{sh}} [1-e^{-t/t_{0}}], & $t < t_{sh}$, \cr
\frac{E_{sh}}{t_{sh}} e^{-t/t_{0}} [e^{t_{sh}/t_{0}}-1], & $t > t_{sh}$,}
\end{equation}
in the case of fixed photospheric radius. Examples of model LCs for those two cases within the context of this type of luminosity input are shown
in Figure 2.

In this type of model
the energy that powers the SN light curve is produced by the diffusion of shock-generated energy through an optically-thick CSM shell of large initial radius. 
Smith \& McCray (2007) adopted a L$\propto r^{2}$ rise for their model of SN~2006gy based on the early portion of the diffusion models of A82.
They did not self-consistently consider the input necessary to drive such a rise. 
This constant power model is related to the ``top hat" magnetar-input model that is considered in Kasen \& Bildsten (2010) with the exception that
we solve for the general case of large initial radius and the diffusion time is defined somewhat differently (see Appendix A). 
We note that in the case where $R_{0}$ is small, the Kasen \& Bildsten result for small initial radius is recovered.
While not especially realistic, this ``top-hat" model captures the essence of the shell-shock model on both the rise and decline.

\subsubsection{{\it Radioactive decays of $^{56}$Ni and $^{56}$Co}}\label{radif}

Diffusion of the radioactive decay energy of newly synthesized $^{56}$Ni and $^{56}$Co has long been
considered the source for powering the LCs of SN Ia, SN Ib/c and some SN II. 
A79, A80, A82 were the first to analytically solve this problem and later 
works by Valenti et al. (2008)
and Chatzopoulos, Wheeler \& Vinko (2009) provide the generalized mathematical expressions for the output LC including
the contribution due to nickel decay and positron heating (assuming the same deposition function) as well as
the effects of gamma-ray leakage.
The output luminosity in this case is found to be:
\begin{eqnarray}
L(t)=\frac{2 M_{Ni}}{t_{d}}e^{-[\frac{t^{2}}{t_{d}^{2}}+\frac{2R_{0}t}{vt_{d}^{2}}]}[(\epsilon_{Ni}-\epsilon_{Co}) 
\int_0^t [\frac{R_{0}}{vt_{d}}+\frac{t'}{t_{d}}]e^{[\frac{t'^{2}}{t_{d}^{2}}+\frac{2R_{0}t'}{vt_{d}^{2}}]}e^{-t'/t_{Ni}}dt'
\nonumber \\
+ \epsilon_{Co} \int_0^t [\frac{R_{0}}{vt_{d}}+\frac{t'}{t_{d}}]e^{[\frac{t'^{2}}{t_{d}^{2}}+\frac{2R_{0}t'}{vt_{d}^{2}}]}
e^{-t'/t_{Co}}dt'](1-e^{-At^{-2}}),
\end{eqnarray}
here $R_{0}$ is the initial radius of the progenitor, $M_{Ni}$ is the initial nickel mass, $t_{Ni} =$~8.8 days,
$t_{Co} =$~111.3 days,
$\epsilon_{Ni} = 3.9 \times 10^{10}$ erg~$s^{-1}$~$g^{-1}$
and $\epsilon_{Co} = 6.8 \times 10^{9}$ erg~$s^{-1}$~$g^{-1}$ are the energy generation rates
due to Ni and Co decays respectively (Valenti et al. 2008). The factor
$(1-e^{-At^{-2}})$ accounts for the gamma-ray leakage, where large $A$
means that practically all gamma rays and positrons are trapped. The gamma-ray optical depth of 
the ejecta is taken to be $\tau_{\gamma} = \kappa_{\gamma} \rho R = At^{-2}$, where $\kappa_{\gamma}$ is
the gamma-ray opacity of the SN ejecta.

Within this model, for which the time-scales of the power inputs are known, 
the following expression for the mass of the SN ejecta can be obtained:
\begin{equation}
M_{ej}=\frac{\beta c R(0) t_{0}}{\kappa} = \frac{3}{10} \frac{\beta c}{\kappa} v t_{d}^{2},
\end{equation}
where $M_{ej}$ is the mass of the SN ejecta, and $\beta$ is an integration constant equal to about 13.8. 
This equation for the diffusion mass (the mass of the SN ejecta in this case) holds for both homologously expanding matter
and fixed photosphere, but only under the assumption of a centrally located power input. If the power input source moves within the diffusion mass,
$t_{0}$ is itself time-dependent and the diffusion mass also changes with time, depending on the position of the source
relative to the that of the photosphere.
Although this model can provide good formal fits to observed LCs of some SLSNe, it cannot be adopted as a general explanation due
to the fact that in many cases the derived $M_{Ni}$ is greater than $M_{ej}$ for reasonable choices of $\kappa$.

The usual assumption of a radioactive decay diffusion model is homologous
expansion of the SN ejecta. We alter this criterion and
consider also the solution for a LC powered by the radioactive decays of $^{56}$Ni and $^{56}$Co, but for the case of fixed photospheric
radius. This may be the case for some SN IIn progenitors that are surrounded by a dense extended optically-thick CSM envelope.
In such case, the photosphere of the relative diffusion mass is coincident with the photosphere of the optically-thick CSM which,
in principle, does not expand homologously or moves with a velocity small compared to the velocity of the SN ejecta.
Although those types of events are primarily powered by shock deposited energy there may still be some contribution from radioactive decay.
In this case, we make use of Equation 4 for a radioactive $^{56}$Ni and $^{56}$Co decay input and we arrive at the following solution:
\begin{eqnarray}
L(t)=\frac{M_{Ni}}{t_{0}}e^{-\frac{t}{t_{0}}}[(\epsilon_{Ni}-\epsilon_{Co}) 
\int_0^t \frac{1}{t_{0}} e^{\frac{t'}{t_{0}}}e^{-t'/t_{Ni}}dt'
\nonumber \\
+ \epsilon_{Co} \int_0^t \frac{1}{t_{0}}e^{-\frac{t'}{t_{0}}}
e^{t'/t_{Co}}dt'](1-e^{-At^{-2}}).
\end{eqnarray}

\subsubsection{{\it Magnetar spin-down}}\label{magdiff}

Recently, Kasen \& Bildsten (2010) and Woosley (2010) considered models of SLSN LCs powered by the spin-down of a young
magnetar (see also Ostriker \& Gunn 1971; Arnett \& Fu 1979; Maeda et al. 2007). 
In such a model, the energy input by the magnetar is given by the dipole spin-down formula:
\begin{equation}
L_{inp}(t) = \frac{E_{p}}{t_{p}} \frac{l-1}{(1+t/t_{p})^{l}},
\end{equation}
where $E_{p}$ is the initial magnetar rotational energy, $t_{p}$ is the characteristic time scale for spin-down 
that depends on the strength of the magnetic field and $l=$~2 for a magnetic dipole. 
For a fiducial moment of inertia, the initial
period of the magnetar in units of 10~ms is given by $P_{10} = (2 \times 10^{50}$erg/s /$E_{p})^{0.5}$. 
The magnetic field of the magnetar can be estimated from $P_{10}$ and $t_{p}$ as $B_{14} = (1.3 P_{10}^{2}/t_{p,yr})^{0.5}$, 
where $B_{14}$ is the magnetic field in units of $10^{14}$~G and $t_{p,yr}$ is the characteristic time scale for spin-down 
in units of years. Incorporating the magnetar spin-down deposition function presented here into Equation 3 and including
the effects of large initial radius, we arrive at the following solution:
\begin{equation}
L(t)=\frac{2 E_{p}}{t_{p}} e^{-[\frac{t^{2}}{t_{d}^{2}}+\frac{R_{0}t}{vt_{d}^{2}}]}
\int_{0}^{x} e^{[z^{2}+\frac{R_{0}z}{vt_{d}}]} \left[\frac{R_{0}}{vt_{d}}+z\right] \frac{1}{(1+yz)^{2}}dz,
\end{equation}
where $x = t/t_{d}$ and $y = t_{d}/t_{p}$ with $t_{d}$ again being an ``effective" diffusion time.
As was the case for the radioactive decay diffusion model, the mass of the SN ejecta for this model is also
given by Equation 10. 

Figure 3 shows an example of a radioactive decay diffusion model LC (solid black curve) compared with a magnetar spin-down model LC (solid red curve)
where the parameters have been chosen so that the output LCs have approximately the same peak luminosity.
The parameters that were used for the models presented in Figure 3 are
$M_{Ni} =$~10~$M_{\odot}$,  $E_{p} =$~$1.37 \times 10^{51}$~erg, $t_{p}$ =~30~days and
for the homologously expanding case for $t_{0} =$~40~days, 
$R_{0} = 10^{14}$~cm and $v =$~10,000~km~s$^{-1}$.
Whereas the decay times of $^{56}$Ni and $^{56}$Co are known experimentally,
the magnetar spin-down model contains two adjustable time-scales, $t_{p}$ and $t_{d}$, instead of just $t_{d}$
as is the case for the radioactive decay diffusion model. The magnetar model therefore provides more freedom for fitting
to observed SN LCs.

\subsection{{\it Shock heating from CSM-ejecta interaction}}\label{csmint}

SN IIn and the recently discovered SLSNe show a variety of LC characteristics in terms of maximum luminosity, duration,
shape and decline rate. Motivated by that, we attempt to use the ejecta-CSM interaction scenario as introduced by
Chevalier (1982) and Chevalier \& Fransson (1994) coupled with diffusion as treated by A80, 82 in order to obtain model
LCs for this type of events.

\subsubsection{{\it Forward and reverse shock luminosity from ejecta-CSM interaction}}\label{csmthin}

We adopt the scenario introduced by Chevalier \& Fransson (1994) in which the progenitor star is embedded
in a CSM shell described by a power-law density profile $\rho_{CSM}=q r^{-s}$, where $\rho_{CSM}$ is the density of the CSM medium,
$q$ is a scaling constant and $s$ the power-law exponent for the CSM density profile. 
In general, $q=\rho_{CSM,1} r_{1}^{s}$, where $\rho_{CSM,1}$ is the density of the CSM shell at $r=r_{1}$. We use as a fiducial value
for $r_{1} = R_{p}$ where $R_{p}$ the radius of the progenitor star. Thus we set the density scale of the CSM, $\rho_{CSM,1}$, immediately
outside the stellar envelope.
For $s=2$, a steady-wind CSM model is recovered, where
$q=\dot{M}/(4\pi v_{w})$ with $\dot{M}$ the pre-SN wind mass-loss rate and $v_{w}$ the pre-SN wind velocity. Deviations
from $s =$~2 may indicate a different pre-SN mass-loss history resulting in CSM clumps or shells. Values of $s$ close to zero
may indicate the presence of CSM ``bubbles" or shells formed by strong stellar winds as proposed for the CSM environments
around Wolf-Rayet stars (Chevalier \& Liang 1989; Dwarkadas 2011). 
The density profile of the SN ejecta is taken to have double power-law profile where the outer ejecta have 
a power-law density profile, $\rho_{SN}=g^{n} t^{n-3}r^{-n}$,
where $g^{n}$ is a scaling parameter for the ejecta density profile, 
$g^{n} = 1/(4 \pi (\delta-n))[2(5-\delta)(n-5)E_{SN}]^{(n-3)/2}/[(3-\delta)(n-3)M_{ej}]^{(n-5)/2}$,
$n$ is the power-law exponent of the outer component, and $\delta$ is the slope of the inner density profile of the ejecta 
(values of $\delta =$~0, 2 are typical), $E_{SN}$ is the total SN energy and
$M_{ej}$ is the total SN ejecta mass.
This assumption for the evolution and the density profile of the SN ejecta is supported by realistic numerical calculations.
The parameter $n$ varies depending on the nature of the progenitor and the presence or absence of a convection zone in the outer parts
of the star. A fiducial value for $n$ is 11.7 corresponding to the case for red supergiant progenitors (Matzner \& McKee 1999) whereas
lower values of $n$ correspond to the envelopes of more compact progenitors.

The interaction between those two media, the SN ejecta and the CSM, each with power-law density profiles results in a 
forward/circumstellar and a reverse/ejecta shock,
the dynamics of which are described by self-similar solutions presented in Chevalier \& Fransson (1994). We use those similarity solutions to derive
the following expression for the luminosity that is produced from this process (neglecting the second, initial value, ``fireball" term
in Equation 3; see Appendix B for the full derivation):
\begin{eqnarray}
L_{inp}(t)=\frac{2 \pi}{(n-s)^{3}} g^{n\frac{5-s}{n-s}} q^{\frac{n-5}{n-s}} (n-3)^{2} (n-5) \beta_{F}^{5-s} A^{\frac{5-s}{n-s}}
(t+t_{i})^{\frac{2n+6s-ns-15}{n-s}} \theta(t_{FS,*}-t)+
\nonumber \\
2 \pi (\frac{A g^{n}}{q})^{\frac{5-n}{n-s}} \beta_{R}^{5-n} g^{n} (\frac{3-s}{n-s})^{3} 
(t+t_{i})^{\frac{2n+6s-ns-15}{n-s}} \theta(t_{RS,*}-t),
\end{eqnarray}
where $\beta_{F}$, $\beta_{R}$ and $A$ are constants that depend on the values of $n$ and $s$ and, for a variety of values, are
given in Table 1 of Chevalier (1982), $\theta(t_{FS,*}-t)$, $\theta(t_{RS,*}-t)$ denote the Heaviside step function that
controls the termination of the forward and reverse shock respectively ($t_{FS,*}$ and $t_{RS,*}$ are the termination time scales for
the two shocks) and $t_{i} \simeq R_{p}/v_{SN}$ is the initial time of the
CSM interaction that sets the initial value for the luminosity produced by shocks where $v_{SN} = [10(n-5)E_{SN}/3(n-3)M_{ej}]^{\frac{1}{2}}/x_{0}$ 
is the characteristic velocity of the SN ejecta, where $x_{0}=r_{0}(t)/R_{SN}(t)$ is the dimensionless radius
of the break in the SN ejecta density profile from the inner flat component (described by $\delta$) to the outer, steeper component
(described by $n$) which is at radius $r_{0}(t)$.
The first and the second terms in Equation 14 refer to the forward and reverse
shock luminosity input, respectively. The forward shock termination time-scale, $t_{FS,*}$, is given by the following expression, assuming
that the input from the forward shock terminates when all the available CSM has been swept up:
\begin{equation}
t_{FS,*}=|\frac{(3-s)q^{(3-n)/n-s)} [Ag^{n}]^{(s-3)/(n-s)}}{4 \pi \beta_{F}^{3-s}}|^{\frac{n-s}{(n-3)(3-s)}} M_{CSM}^{\frac{n-s}{(n-3)(3-s)}},
\end{equation}
where $M_{CSM}$ is the total mass of the CSM.  
Once the forward shock breaks out from the optically-thick part of the CSM envelope, the luminosity deposition
from it is primarily in the UV/X-ray region of the spectrum, not the optical. The time $t_{FS,*}$ can represent the breakout time if $M_{CSM}$ is taken
to be the mass of the optically-thick CSM, rather than the total mass. We adopt this assumption below.
Assuming that the reverse shock terminates when all the ejecta are swept-up,
the reverse shock termination time-scale, $t_{RS,*}$, is given by the 
following expression:
\begin{equation}
t_{RS,*}=[\frac{v_{SN}}{\beta_{R} (Ag^{n}/q)^{\frac{1}{n-s}}} (1-\frac{(3-n)M_{ej}}{4\pi v_{SN}^{3-n} g^{n}})^{\frac{1}{3-n}}]^{\frac{n-s}{s-3}}.
\end{equation}
where $M_{ej}$ is the total ejecta mass.

To verify the analytical result of Equation 14, we compare it to the LC that is calculated
by a two-dimensional simulation of the collision of SN ejecta with a wind CSM component ($s=2$) as presented
by Van Marle et al. (2010). We adopt the same parameters for the CSM and the SN ejecta components as in
their model O01 ($n =$~11.7, $s =$~2, $v_{w} =$~200~km~s$^{-1}$ and $\dot{M} =$~$10^{-4}$~$M_{\odot}$~yr$^{-1}$) and present
the results in Figure 4.
The agreement between our analytical result (red solid curve) and their simulation (filled circles) is remarkable given that they 
have used a different technique to calculate the radiated luminosity in their simulation. 
 
\subsubsection{{\it Ejecta-CSM interaction with diffusion}}\label{csmthick}

The spectrum produced by optically-thin ejecta-CSM interaction,
is expected to be a hard spectrum populated with emission lines produced by forward and reverse shocks (Chevalier \& Fransson 1994; Nymark, Chandra \&
Fransson 2009). 
Spectra with these characteristics have not been seen in SLSNe so far,
so it is natural to assume that the effects of radiative diffusion of the shock-generated luminosity are important
in explaining both the currently observed optical light curves and optical spectra of these events. 
For this reason we develop an analytical model that couples the effects of a CSM shock heat input as derived
in the previous section, with the prescription of radiative diffusion as treated by A80, 82. We will consider
the case of a stationary CSM photosphere for this problem and assume that the photosophere is somewhere in the CSM envelope where
it will not, in principle, be rapidly expanding. 
In these models we ignore ``backwarming" of the photosphere by any shock that has propagated beyond the CSM photosphere.
We include the effects of terminated shock luminosity input.

We start by assuming that the luminosity input from the forward and the reverse shock happens deep within the fixed photosphere of the CSM
and therefore we treat it as centrally located. 
This assumption coupled with the nature of the CSM we explore means that the typical shock crossing time scale,
$R_{sh}/v_{sh}$, is larger than the effective radiation diffusion time scale $t_{d}$. This statement is equivalent to the condition
that the CSM optical depth is smaller than the characteristic optical thickness of a radiation-dominated and radiation-mediated shock, 
$c/v_{sh}$, and thus in the regime considered by
Nakar \& Sari (2010) who calculate shock breakout LCs for a variety of SN progenitors (see also Ensman \& Burrows 1992). 
Chevalier \& Irwin (2011) investigated situations
in the opposite regime, $R_{sh}/v_{sh} < t_{d}$, for a steady state wind ($s =$~2), 
considering the effect of the propagation of a radiation-dominated forward shock into the mass-loss region for the cases
where the characteristic radiation breakout radius ($R_{d}$, as defined in their work) is either smaller or larger 
than the wind termination radius ($R_{w}$). In Chevalier \& Irwin (2011), diffusion occurs when the forward shock reaches $R_{d}$
and $R_{sh}/v_{sh} \sim t_{d}$. Subsequently,
the diffusion would proceed more rapidly than the shock. This is the regime we consider. We note that when the condition $R_{sh}/v_{sh} > t_{d}$
is satisfied, the shock is no longer radiation mediated.
Ofek et al. (2010)  used a related model to explain the LC of the IIn SN PTF~09uj.
 Balberg \& Loeb (2011) presented a similar model, but for a less dense
wind in the context of observational signatures of the UV/X-ray shock breakout LC.

Both the forward and the reverse shocks move through the diffusion mass affecting
the radiation diffusion time and changing the form of the output LC. Accounting for the movement of the sources makes the
resulting PDE unseparable and the problem hard to solve analytically, and we neglect this aspect in the current models. 
We do, however, account for the shock propagation by terminating the shock luminosity input after the specific time scales that are given by Equations 15 and 16. 
We assume that the forward shock input within the optically thick part of the CSM
terminates at a time given by Equation 15 for $M_{CSM} = M_{CSM,th}$, where $M_{CSM,th}$ is
the mass of the optically-thick part of the CSM. This shock termination time is close
to the time of forward shock breakout, $t_{FS,BO}$, which formally occurs when the shock
reaches optical depth $tau = v_{sh}/c$. The mass of the optically-thick CSM is given by:
\begin{equation}
M_{CSM,th}=4 \pi q \int_{R_{p}}^{R_{ph}} r^{2-s} dr,
\end{equation}
and we use optical depth,
\begin{equation}
\tau = \kappa q \int_{R_{CSM}}^{R_{ph}} r^{-s} dr = \frac{2}{3},
\end{equation}
to determine the radius of the photosphere, $R_{ph}$, where $\kappa$ is the optical opacity and
$R_{CSM}$ is the total radius of the CSM determined by the following formula:
\begin{equation}
M_{CSM} = 4 \pi q \int_{R_{p}}^{R_{CSM}} r^{2-s} dr,
\end{equation}
where $M_{CSM}$ is the total mass of the CSM envelope that we use as the basic parameter.
Solving Equation 19 with respect to $R_{CSM}$, Equation 18 for $R_{ph}$ and setting the results back into Equation 17 we obtain
the final result for $M_{CSM,th}$ as an expression that depends on the basic model parameters $M_{CSM}$, $q$, $s$, $R_{p}$ and $\kappa$.
Once the forward shock breaks out, the bolometric luminosity will still invoke the full forward/reverse
shock input given by Equation 14, but the optical luminosity will decline since the bulk of the energy will be emitted in the UV/X-ray region of the spectrum.
A portion of the optically-thin forward shock contribution is expected to be in the optical due to re-radiation and electron scattering (Chevalier \& Fransson 1994),
but we neglect this effect. With this model we can also estimate the time and intensity of the rise of UV/X-ray flux due to forward shock breakout.

Now we can implement the SN ejecta-CSM interaction luminosity input from the forward and the reverse shock given by Equation 14 into the first term of 
Equation 4 that accounts for the diffusion through an optically-thick CSM with a fixed photosphere and obtain the following final expression 
for the output model LC:
\begin{eqnarray}
L(t)=\frac{1}{t_{0}} e^{-\frac{t}{t_{0}}} \int_0^t e^{\frac{t'}{t_{0}}} 
[\frac{2 \pi}{(n-s)^{3}} g^{n\frac{5-s}{n-s}} q^{\frac{n-5}{n-s}} (n-3)^{2} (n-5) \beta_{F}^{5-s} A^{\frac{5-s}{n-s}}
(t'+t_{i})^{\frac{2n+6s-ns-15}{n-s}} 
\nonumber \\
\times \theta(t_{FS,BO}-t') + 2 \pi (\frac{A g^{n}}{q})^{\frac{5-n}{n-s}} \beta_{R}^{5-n} g^{n} (\frac{3-s}{n-s})^{3} 
(t'+t_{i})^{\frac{2n+6s-ns-15}{n-s}} \theta(t_{RS,*}-t')]dt',
\end{eqnarray}
where, in the case of an optically-thick CSM, $t_{0} = \kappa M_{CSM,th} /\beta c R_{ph}$. After the times of the termination of the shock
power inputs, the luminosity decays according to Equation 6 where the initial value of the luminosity is now that for the corresponding
terms in Equation 20 at $t_{FS,BO}$ and $t_{RS,*}$, respectively.

The analytic model described by Equation 20 for
the output LCs for SNe powered by SN ejecta-CSM interaction depends on the properties of the progenitor star ($E_{SN}$, $M_{ej}$, $R_{p}$, 
$\delta$ and $n$) and the properties of the CSM ($s$, $\rho_{csm,1}$, $\kappa$ and $M_{CSM}$). 
The fact that there are many unknown parameters involved in this problem follows from the complex nature of 
CSM interaction due to the large variety of possible CSM environments (winds, shells of any density scale) and the large variety
of possible SN progenitors (red or blue supergiant stars). This natural variety can explain certain
differences in the observed optical LC shape, duration and luminosity of SN IIn. In the same context, the faster post-maximum decline of a SN IIL LC
can be the result of particular conditions in this large parameter space involved in SN ejecta-CSM interaction, as we discuss next.

\subsubsection{{\it Ejecta-CSM interaction and radioactive decay input with diffusion}}\label{csmthickrad}

Here we consider a hybrid model in which the luminosity input is provided by both SN ejecta-CSM interaction and by the radioactive decays of
$^{56}$Ni and $^{56}$Co. To find the combined luminosity input we add the radioactive decay luminosity input to the forward and reverse shock
luminosity input given by Equation 14. Figure 5 shows an example of this hybrid luminosity input for collision of SN ejecta 
($E_{SN} = 1.5 \times 10^{51}$~erg,
$M_{ej} =$~20~$M_{\odot}$, $R_{p} = 10^{14}$~cm, $n =$~12) with an optically-thick massive CSM shell 
($s =$~0, $M_{CSM} =$~1~$M_{\odot}$) in the case where 1~$M_{\odot}$ of $^{56}$Ni
is formed in the explosion, which might be the case for a hypernova or a pair-instability SN (PISN). The 
early behavior of the input is controlled by the relative contribution between the shock inputs and
the radioactive decay, and the very late-time decay rate is equal
to the $^{56}$Co radioactive decay rate. In the cases where CSM interaction is dominant, the early decline of the input simply scales as a power law function of time with the
power being a function of the values of $n$ and $s$ as given by Equation 14.

To obtain a model bolometric LC for this hybrid input that can be used for fits to observed SN LCs we assume that
the radioactive decay deposition takes place within the whole diffusion mass ($M_{ej}+M_{CSM,th}$). We assume the CSM interaction luminosity
input takes place just within $M_{CSM,th}$, since in this context we assume the shocks to be ``frozen" at the interaction region. 
We take the interaction region to be
the interface between the edge of the progenitor star and the CSM envelope and deep within the CSM photosphere.
Therefore, the final hybrid CSM interaction plus radioactive decay LC model has the following form:
\begin{eqnarray}
L(t)=\frac{1}{t_{0}} e^{-\frac{t}{t_{0}}} \int_0^t e^{\frac{t'}{t_{0}}} 
[\frac{2 \pi}{(n-s)^{3}} g^{n\frac{5-s}{n-s}} q^{\frac{n-5}{n-s}} (n-3)^{2} (n-5) \beta_{F}^{5-s} A^{\frac{5-s}{n-s}}(t'+t_{i})^{\frac{2n+6s-ns-15}{n-s}} \theta(t_{FS,BO}-t') 
\nonumber \\
+ 2 \pi (\frac{A g^{n}}{q})^{\frac{5-n}{n-s}} \beta_{R}^{5-n} g^{n} (\frac{3-s}{n-s})^{3} (t'+t_{i})^{\frac{2n+6s-ns-15}{n-s}} \theta(t_{RS,*}-t')]dt'+\frac{1}{t_{0}'} e^{-\frac{t}{t_{0}'}} 
\int_0^t e^{\frac{t'}{t_{0}'}} M_{Ni}
\nonumber \\
\times [(\epsilon_{Ni}-\epsilon_{Co})e^{-t'/t_{Ni}} + \epsilon_{Co}e^{-t'/t_{Co}}]dt',
\end{eqnarray}
where $t_{0}$ and $t_{0}'$ correspond to the diffusion time-scales through $M_{CSM}$ and $M_{ej}+M_{CSM,th}$ respectively.
Once again, the forward and reverse shock terms are replaced by pure diffusion decay terms analogous to Equation 6 when shock input terminates.
Figures 6 and 7 show some model SN LCs that result from SN ejecta-CSM interaction with a shell ($s =$~0) and a steady-state wind ($s =$~2),
respectively. Examples are plotted for a variety of input luminosity combinations so that the effect on the final output LC can be illustrated.
The parameters used for the models plotted are listed in the captions of the figures. Some of the peaks of the resulting model SN LCs are not smooth
due to the fact that we have used a simplistic Heaviside function prescription for the termination of the forward and reverse shock luminosity input.
In reality this termination process will be a smoother function of time, thus making the peak of those LCs smoother, too. 

Note in Figures 6 and 7 that when the shock inputs rise monotonically with time (shell model; Figure 6) the LC on the rise has a monotonically
increasing slope, but when the shock inputs decline monotonically with time (wind model; Figure 7), the LC on the rise has a monotonically decreasing
slope, reminiscent of most observed SNe. The shape of the rising LC is thus a potentially strong constraint on the models.
Given the large number of parameters involved in the hybrid model described by Equation 21, a careful survey of the parameter space is needed in
order to investigate issues of parameter degeneracy and correlation. We will present such an analysis in a follow-up paper that will illustrate 
fits of our model to observed SLSN LCs. From a qualitative perspective, model LCs presented here will generally be brighter and briefer for
higher values of $E_{SN}$, $R_{p}$ and $\rho_{CSM,1}$ and lower values of $M_{CSM}$. Luminosities characteristic of SLSNe are more effectively
produced considering interactions with dense shells ($s =$~0) rather than with steady-state winds ($s =$~2). 
Models with constant density shells tend to be more sensitive to $R_{p}$ than those with winds because of the effect of the forward and reverse shocks.
Interactions with steady-state
winds are capable of producing LCs reminiscent of some normal luminosity SNe IIn, as we will show in the follow-up paper.

We now use the hybrid model presented in Equation 21 to investigate the variety of LC shapes that we can obtain from that model and discuss implications
for the SN II classification scheme. As can be seen in Figure 8, a hybrid CSM interaction plus radioactive decay input in which CSM interaction
dominates can produce a faster post-maximum decline rate that is consistent with the observed 
rapid post-maximum decline rate (``linear" in the logarithmic scale) of SN IIL. This is due to the termination of the dominant luminosity input
(at $t = t_{RS,*}$ in the case illustrated, for which the reverse shock luminosity dominates) that will lead to a fast post-maximum decline dominated by cooling
and diffusion of previously deposited shock luminosity.
Furthermore, the differences seen between SN IIL and SN IIn in their spectra, namely the absence of P Cygni absorption components in the Balmer lines
in the latter, maybe due to differences in the optical depth of the CSM involved: A higher optical depth would obscure the effects of the underlying
SN photosphere expansion thus leading to significant CSM interaction and the presence of narrow Balmer emission lines without a detectable broad absorption component, 
whereas a lower optical depth in the CSM allows for the underlying SN expansion to be seen and the effects of the CSM interaction
are mainly imprinted in the LC of the event. Therefore, the SN IIL class may represent a subset attributable
to a particular restricted range in the large number of parameters involved in the more generic SN ejecta-CSM interaction scenario and thus
SN IIL may be a subclass of SN IIn. 

The generic model presented here might also be used to explain the existence of approximately symmetric SN LCs like the ones observed for the 
transients SCP06F6 (Barbary et al. 2009), PTF09cwl, PTF09cnd and PTF09atu (Quimby et al. 2011). An example of such a ``symmetric" LC
around the peak is also seen in Figure 8 when CSM interaction and radioactive decay are comparable. This topic will be discussed in more
detail in the follow-up paper.

This hybrid model also allows us to predict when the forward shock will break out of the CSM photosphere based on the fit to 
the optical light curve. The shock breakout will be followed by an X-ray/UV burst
of radiation. The model may also be used to estimate the intensity of that radiation and the subsequent UV/X-ray LC. In Figure 9 we illustrate this 
by plotting the optical and UV/X-ray output LCs for a choice of parameters given in the captions for the case of SN ejecta interaction with a constant 
density shell ($s =$~0; left panel) and a wind ($s =$~2; right panel).

\subsubsection{{\it Comparison with results from radiation hydrodynamics modeling}}\label{numericalcomp}

The various approximations associated with our analytic hybrid SN LC model and its limitations do not allow us to use it to address more
specific emission characteristics like radio emission, thermalization of the radiation, the ionization state, the change in the opacity of the gas, 
and the evolution of the photosphere properties. Those aspects can be addressed only via numerical radiation hydrodynamics simulations and are beyond
the scope of this project. We can, however, address the limitations and uncertainties of our model bolometric 
SN LCs by benchmarking our
results against existing numerical calculations for the same initial conditions. To do so, we have used the results from three sets of numerical simulations:
the Woosley, Blinnikov \& Heger (2007) (hereafter WBH07) simulation of the LCs produced by a pulsational pair-instability supernova (PPISN), 
the Moriya et al. (2011)
simulations of the interaction between SN ejecta from RSG progenitors with optically-thick CSM winds and the Chugai et al. (2004) simulation of
the collision of SN ejecta with a 0.4~$M_{\odot}$~CSM with density profile $\rho_{CSM} \sim r^{-1}$ in order to reproduce the LC of SN~1994W. 
We note that in our treatment we neglect the 
initial offset that the LCs will have in the time axis due to the fact that the first light will emerge after a diffusion time-scale. 
This offset is shown in the simulations of WBH07 and Moriya et al. (2011) and we can also explicitly calculate it in each case.

WBH07 considered the evolution of a 110~$M_{\odot}$ star that ends its life with an oxygen core of 
$\sim$~50~$M_{\odot}$ that is in
the domain for which the core is unstable due to electron-positron pair production. The collapse and subsequent pulse is not strong enough to disrupt the 
whole star,
but only to eject 24.5~$M_{\odot}$ of its outer parts. The remainder of the star (with radius $\sim 10^{14}$~cm) contracts, 
encountering pair-instability again after 6.8 years that leads to a second, more energetic pulse ($6 \times 10^{50}$~erg) that ejects 5.1~$M_{\odot}$. 
The ejecta from this second pulse then collides with the slower moving ejecta from the first pulse producing a luminous SN-like output LC that was calculated using
the code STELLA (Blinnikov \& Bartunov 1993; Blinnikov et al. 1998; Blinnikov et al. 2006) and presented in supplementary Figure 6 of WBH07. The authors
consider this PPISN phenomenon as an explanation for the SLSN~2006gy.

To reproduce the WBH07 result, we use our hybrid model, presented by Equation 21 above, for the same parameters
($E_{SN} =$~$6 \times 10^{50}$~erg, $R_{p} =$~$10^{14}$~cm, $M_{ej} =$~5.1~$M_{\odot}$, $M_{CSM} =$~24.5~$M_{\odot}$, $M_{Ni} =$~0~$M_{\odot}$). 
The parameter $\rho_{CSM,1}$ is not explicitly given in WBH07 however we can estimate it, assuming steady-state mass loss. In this case
$\rho_{CSM,1} = \dot{M}/(4\pi v_{w} R_{p}^2)$ and $\dot{M} \simeq$~(24.5~$M_{\odot})$/(6.8 years), so for $v_{w} \simeq$~1,000~$km$~$s^{-1}$ 
$\rho_{CSM,1} \simeq$~$1.8 \times 10^{-11}$~g~cm$^{-3}$. This value is close to the dense shell values seen in supplementary Figures 10 and 11 of WBH07.
The density profiles of the SN ejecta (second pulse) and the CSM (ejecta from first pulse) are also uncertain and not necessarily described
by power laws in the context of a PPISN. 
For the purposes of our comparison, we will assume that the SN ejecta from the second pulse has a SN-like power law density profile with
slope $n =$~12 and for the CSM we will consider two cases: one for a constant density shell ($s =$~0) and one for a steady state wind ($s =$~2).
We also assume in our model that the photosphere is within a stationary CSM while in the model of WBH07
the first shell which comprises the CSM is expanding homologously at a speed of 100-1,000~$km$~$s^{-1}$. 
The results of our comparison are shown in Figure 10. The solid black and red curves correspond to the bolometric LCs for the original ejecta velocity and a doubled
ejecta velocity, respectively, and the dotted black and red curves correspond to the UBVRI LCs as calculated by WBH07. The dashed curves show our
analytic LCs for $s =$~0 (green curve), for $s =$~2 (blue curve) and for $s =$~2 but for the values of $M_{ej}$ and $M_{CSM}$ 
adjusted in order to provide a better fit to the results from the simulations (yellow curve). 
As can be seen, given the uncertainties of the density profiles
and the simplifying assumptions of our model (centrally located power input and a fixed photosphere in the CSM) the two models
agree in terms of rise time to maximum light and peak luminosity. The post-maximum decline rates of our models are similar
to those found by WBH07, but of higher luminosity. A somewhat better agreement is found if we decrease $M_{ej}$ and $M_{CSM}$ by
about 50 \%, giving some indication of the uncertainty in our ability to estimate parameters.

Bolometric and color LCs of SNe produced by the interaction of SN ejecta (from a RSG progenitor) with a CSM that is a steady state wind ($s =$~2)
or with a $s =$~1.5 density profile slope were presented by Moriya et al. (2011) as results of 1-D numerical radiation hydrodynamics
calculations. We compare the output LCs from our analytic hybrid model to those presented by Moriya et al. (2011) for four of their models:
s13hw2r20m2e3 ($E_{SN} =$~$3 \times 10^{51}$~erg, $R_{p} =$~$5 \times 10^{13}$~cm, $M_{ej} =$~8~$M_{\odot}$, 
$M_{CSM} =$~0.65~$M_{\odot}$, $\rho_{CSM,1} =$~$2 \times 10^{-11}$~g~cm$^{-3}$),
s15hw2r10m3e3 ($E_{SN} =$~$3 \times 10^{51}$~erg, $R_{p} =$~$5 \times 10^{13}$~cm, $M_{ej} =$~10~$M_{\odot}$, 
$M_{CSM} =$~0.031~$M_{\odot}$, $\rho_{CSM,1} =$~$2 \times 10^{-12}$~g~cm$^{-3}$),
s15hw2r20m2e1 ($E_{SN} =$~$10^{51}$~erg,          $R_{p} =$~$5 \times 10^{13}$~cm, $M_{ej} =$~10~$M_{\odot}$, 
$M_{CSM} =$~0.65~$M_{\odot}$, $\rho_{CSM,1} =$~$2 \times 10^{-11}$~g~cm$^{-3}$) and 
s15hw2r20m3e3 ($E_{SN} =$~$3 \times 10^{51}$~erg, $R_{p} =$~$5 \times 10^{13}$~cm, $M_{ej} =$~10~$M_{\odot}$, 
$M_{CSM} =$~0.065~$M_{\odot}$, $\rho_{CSM,1} =$~$2 \times 10^{-12}$~g~cm$^{-3}$). 
The details for the parameters used for these models are given in Table 2 of Moriya et al. (2011). 
Figure 11 shows the result of this comparison. The solid curves give the model LCs from the Moriya et al. (2011) simulations
and the dashed curves our analytical model LCs. Our models again produce the rise time to maximum light
and peak luminosity as well as the width of the main diffusion curve. Our post-maximum decline rates are again consistently similar to or slower than the ones
found by the simulations. 

Chugai et al. (2004) sucessfully reproduced the plateau-like LC of the SN IIn SN~1994W with radiation hydrodynamics simulations that involved
contribution from both the radioactive decay of $^{56}$Ni and interaction with a dense CSM. More specifically they computed
three models, of which their sn94w58 model provided the best fit to the data. This model considered the explosion of a progenitor with a very extended
radius ($R_{p} =$~$1.4 \times 10^{15}$~cm; their wind termination radius is several times larger) 
with energy $E_{SN} =$~$1.5 \times 10^{51}$~erg and $M_{ej} =$~7~$M_{\odot}$ that produced 0.015~$M_{\odot}$ of 
radioactive $^{56}$Ni within a 0.4~$M_{\odot}$~CSM with density profile slope $s =$~1. Given those parameters the scale density at the base of the CSM 
is $\rho_{CSM,1} =$~$1.2 \times 10^{-14}$~g~cm$^{-3}$. 
Using those parameters, we plot our analytical LC model (dotted curve) and compare it to the Chugai et al. (2004) numerical
V-band (solid green curve) and U-band (solid blue curve) LCs in Figure 12. 
We again see that the overall agreement between the models is very good. Our analytical model reproduces the plateau of SN~1994W 
in terms of duration and luminosity. We also observe that the analytical LC model is closer to the U-band LC numerical model; it
reproduces the two luminosity breaks of the LC as a result of forward (early break) and reverse (later break) shock input termination.

The comparison between our analytic model LCs and radiation hydro LCs for the same conditions presented here illustrates the uncertainties
involved. The fact that we consider the forward and reverse shock power input to be centrally located and the photosphere to be 
fixed in the CSM in order to obtain a separable PDE leads to an overestimate of the diffusion time scale in our models because 
the relative shock and photosphere dynamics in reality will lead to an ever decreasing optical depth between the forward shock
and the photosphere in the CSM.
This will have the effect of producing 
a LC with a slower post-maximum decline rate in our models, which is what we see in Figures 10 and 11. Consequently,
the overestimate of the diffusion time scales means we must employ smaller diffusion masses ($M_{CSM}$ and $M_{ej}$) to better fit a given LC.
For a given LC, we will tend to underestimate the masses involved. 
This is illustrated by our adjusted $s =$~2 model to better reproduce the result of WBH07. The optical depth and the diffusion characteristics 
will also be affected by proper treatment of the acceleration of the CSM by the precursor wave, the ionization state of the 
CSM, and the position of the ionization front taking into account the decline in optical opacity beyond it, effects considered by
Moriya et al. (2011).

Our comparison with the models of Moriya et al. (2011) are a particularly stringent test, since the CSM in
their models was of rather low mass and our models are designed to represent the case of especially massive CSM shells. 
The breaks in luminosity seen in the two classes of models are of somewhat different, but closely, related nature. In our case, the first break in 
luminosity occurs when the forward shock breaks out of the optically-thick CSM or when the reverse shock has swept up all 
the available mass of SN ejecta. For Moriya et al. (2011), these breaks are the result of the photosphere receding within the 
interaction region; the late time radiation is due to ``left-over" thermal emission within the ejecta. In our models, the
shock moves out to the photosphere; in the models of Moriya et al. (2011) the photosphere moves inward to meet the forward
shock, thanks in large part to the acceleration of the CSM by the radiation flow. For substantially more massive CSM the radiative acceleration will be less,
and we would expect our models and those of Moriya et al. to converge to greater similarity. As noted above, we get quite good agreement with the
results of Chugai et al. (2004) for models with a more massive CSM.

Although the physics involved is accounted for in a more accurate way in the numerical simulations, our simple analytic results
reproduce the basic LC features. We understand the sign of the effect of neglecting the decrease in optical depth in front of the forward shock.
Our models may thus be used as a first step in fitting observed SN LCs to get a basic understanding of the parameters involved before
proceeding to more expensive numerical simulations.

\section{APPLICATION TO THE OPTICAL LC OF SLSN~2006gy}\label{06gyfit}

SLSN~2006gy stirred a lot of discussion among the SN community. It was discovered by the ROTSE-IIIb telescope of the Texas Supernova Search project 
(Smith et al. 2007), which obtained unfiltered photometry over the course of $\sim$~200d providing us with a well-constrained LC and
explosion date. We allow the explosion date to vary in a limited range in our fitting process
that will be discussed below. The very late-time decline rate of the LC of SLSN~2006gy is reported to be consistent 
with the decline of $^{56}$Co (Smith et al. 2007).
 A rich database of optical spectra were obtained for SLSN~2006gy (Smith et al. 2007, 2008, 2010) that provides
an extensive record of its spectral evolution. SLSN~2006gy showed strong Balmer emission features with their narrow components
associated with P Cygni absorption indicative of photospheric expansion. The H$\alpha$ line profile evolved throughout the course of the LC
of SLSN~2006gy showing an evolution that is marked by three phases described in Smith et al. (2010). The full width at half-maximum (FWHM) of H$\alpha$
around maximum light reveals characteristic velocities of $\sim$~4,000~km~s$^{-1}$. We note, however, that this velocity information may not directly
correspond to the bulk kinematic motion of the SLSN ejecta. Smith et al. (2010) discuss the possibilities of line broadening due to electron scattering
that yields information about the scattering optical depth and thermal motion of the electrons, but not about a true physical bulk expansion velocity 
associated with SLSN~2006gy.
We will nevertheless use the value of $\sim$~4,000~km~s$^{-1}$ as a fiducial velocity for our model fitting purposes. Spectra of the host galaxy 
of SLSN~2006gy determined
the redshift of the SLSN to be $z =$~0.074, which means that the absolute visual peak magnitude of the event reached $\sim$~-22~m, making SLSN~2006gy
one of the brightest explosions ever discovered. Based on the information given by the optical LC and the spectra, SLSN~2006gy was
classified as a SN IIn event (Smith et al. 2007).

Various models have been discussed to explain the nature of SLSN~2006gy. Smith \& McCray (2007)
considered a shell-shock diffusion model for which the luminosity output around maximum light is reproduced by shock heating due to the interaction
between the SN ejecta and a dense CSM shell. 
Their simple model provided a decent fit for the decline of the LC, but failed to properly account for the rise of the LC since
they considered an instantaneous shock input at maximum light. They also considered a contribution from the radioactive decay of $^{56}$Co to account
for the LC at very late times and estimated a $^{56}$Ni mass of about $\sim$~8~$M_{\odot}$. 
A pair-instability SN (PISN) scenario was discussed for SLSN~2006gy in Smith et al. (2007). This model has difficulties in accounting for
the large discrepancy between the total ejected and nickel mass and it turns out that it is not a good fit to the data, as we will show below.
Other models considered interaction with an extensive CSM envelope (Smith et al. 2007) or CS clouds (Agnoletto et al. 2009). A popular picture today
is that SLSN~2006gy resulted from the explosion of a massive LBV-type star within a dense CSM envelope, 
reminiscent to that of $\eta$~Car (Smith et al. 2007). Another model
for SLSN~2006gy was considered by WBH07 in which the LC of the SN is due to interaction between shells ejected as a result
of the PPISN process (\S3.3.4).
We also note that another, exotic model, a quark-nova explosion has been considered for SLSN~2006gy and other SLSNe (Ouyed et al. 2010).

Given that so far we are lacking a self-consistent LC model for SLSN~2006gy that reproduces the whole LC and accounts for its spectral characteristics,
we consider the ROTSE LC of SLSN~2006gy converted to a pseudo-bolometric LC to test the hybrid 
model discussed in this paper. We assume the bolometric correction BC=0 due to the fact that 
Smith et al. (2007) do not provide such an estimate
since there are not adequate simultaneous multi-band photometric observations throughout the course of the LC of the event. 
Therefore we accept the lack of a true observed bolometric LC of SLSN~2006gy as one more uncertainty in our fit. We also adopt $E(B-V) =$~0.72~mag
yielding R-band extinction $A_{R} =$~1.68~mag (Smith et al. 2007).
We fit Equation 21 to the pseudo-bolometric LC of SLSN~2006gy and present our result in Figure 13. 
Ideally the fit would be done by a chi-square minimization technique (Chatzopoulos, Wheeler \& Vinko 2009), but here we have chosen parameters
by hand to illustrate the capacity of the model to represent the observations.
The model illustrated was obtained for the following
choice of parameters: $\delta =$~0, $n =$~12, $s =$~0, $E_{SN} = 4.4 \times 10^{51}$~erg, $M_{ej} =$~40~$M_{\odot}$, 
$M_{CSM} =$~5~$M_{\odot}$ ($M_{CSM,th} =$~4.9~$M_{\odot}$), $R_{p} =$~$5 \times 10^{14}$~cm, 
$\rho_{CSM,1} =$~$1.5 \times 10^{-13}$~g~cm$^{-3}$, 
$M_{Ni} =$~2~$M_{\odot}$. For these parameters we estimate the radius of the photosphere to be $R_{ph} =$~$2.5 \times 10^{15}$~cm
and the optical depth of the CSM $\tau_{CSM} \sim$~120, which is consistent with the fact that the photosphere is within the optically
thick shell. As argued in the previous section 
where we presented a comparison with the radiation hydrodynamics results of Woosley, Blinnikov \& Heger (2007) for the LC of SLSN~2006gy, 
our masses may be underestimated by $\sim$~50 \%. Given the maximum luminosity
of SLSN~2006gy ($L_{max} =$~$1.2 \times 10^{44}$~erg~s$^{-1}$) and the radius of the photosphere we derived, we estimate a black-body
temperature of $T_{BB} \simeq$~13,000K. This value for $T_{BB}$ is consistent with the value $T_{BB} =$~11,000~-~12,000K 
that Smith et al. (2007) derived from low-resolution spectra taken close to maximum light.
We note, however, that black-body emission generally provides a poor fit to the SEDs of
SNe IIn due to flux dilution and line blanketing and to the fact that other emission
mechanisms are dominant (see for example Dessart et al. 2009 on SN 1994W, Miller et al. 2010 on SN 2008iy; 
Chatzopoulos et al. 2011 on SN 2008am).

The fit to the LC of SLSN~2006gy given in Figure 13 was based on a constant density shell. 
To illustrate the utility of our models to limit parameters, we note that it was
diffcult to provide an equally satisfactory fit based on a finite mass, ``steady-state" wind, with $s =$~2.  The CSM luminosity input is generally too small to
account for the observed luminosities even for large values of the CSM scale density, $\rho_{CSM,1}$. 
Extraordinary values for $\rho_{CSM,1}$ (10$^{-9}$, 10$^{-8}$, 10$^{-7}$~g~cm$^{-3}$) imply a
wind that starts at the stellar photosphere, but with a higher density than the outer envelope of the star. This condition is unphysical and implies extraordinary
mass-loss rates ($>$100 solar masses per year). Even with extraordinary values for $\rho_{CSM,1}$ 
there are fitting issues because the input for the forward and reverse shocks in the $s =$~2 case declines rapidly monotonically, thus forcing 
the rise to maximum to be very rapid. Figures 7 and 9 (second panel) illustrate $s =$~2 cases for which the rise time is less than 15-20 days. 
For $s =$~2, it is very difficult to produce a longer rise time without, for instance, having a wind mass considerably in excess of the ejecta mass and an 
overall poor fit in detail. Wind models that roughly give the correct maximum
luminosity also give a continuously-declining CSM shock input that produces a very slow decline of the output luminosity until shock termination occurs that does
not agree with the observations. We thus find that while the models do have a large number of parameters, the observations considerably constrain the model
parameters in practice.

We also plot in Figure 13 (dashed curve) the expected UV/X-ray LC due to the forward shock propagation in the optically-thin part of the CSM 
based on the parameters determined by the fit to the optical LC. We see that the forward-shock luminosity, and associated UV/X-ray burst,  
is terminated at about 54 days after explosion in the rest-frame. 
This may be consistent with the low X-ray flux detected by {\it Chandra} at about 54 rest-frame days after explosion as reported by Smith et al. (2007). 
Once the forward shock terminates upon reaching the outer edge of the shell, 
the UV/X-ray luminosity falls dramatically. This result also implies that the UV/X-ray burst
prior to termination of the forward shock would have been bright enough to be detectable.

We conclude that interaction of typical SN ejecta with a massive optically-thick CSM shell plus a contribution from 2~$M_{\odot}$ of radioactive Ni
seems to reproduce the observed LC of SLSN~2006gy. In this context, the progenitor of SLSN~2006gy was probably a massive star that underwent significant
and episodic mass loss. The massive CSM shell could have been the result of either LBV-type ejection, or a bubble due to interaction of stellar winds
of previous epochs or the result of a PPISN. We also note that our hybrid CSM interaction plus radioactive decay model fit for SLSN~2006gy is similar
to the idea presented by Agnoletto et al. (2009) in the sense that they also used arguments for combined CSM interaction (with CS clumps 
in their decription) and radioactive decay of less nickel than previously implied (3~$M_{\odot}$).

\section{DISCUSSION AND CONCLUSIONS}\label{disc}

 We derived semi-analytical models for the LCs of SN II by incorporating the effects of diffusion as treated by A80, 82 with a variety
of inputs that are considered candidate SN LC powering mechanisms (radioactive decays of $^{56}$Ni and $^{56}$Co, magnetar spin-down and ejecta-CSM
interaction). We found solutions for the cases of a fixed (in radius) CSM photosphere and homologously expanding matter, including the effects
of terminated power input in the case of CS shock heating.  

 One implication from our results is that the principal factor that gives rise to the observed diversity in the shape, duration and luminosity
of SN IIn 
is the presence or absence of a CSM enviroment and the physical properties that are associated with it; mainly its optical
depth, characteristic density and density profile. SN IIP, SN Ib and SN Ic are all the result of core-collapse explosions
with the difference that in SN Ib and SN Ic events the H and He envelope of the progenitor star is lost and in SN IIP the large outer H envelope
is retained. The CSM enviroment around these events is probably of very low optical depth, so that the effects
of ejecta-CSM interaction are negligible and the only mechanism that gives rise to the optical luminosity output are shock heating of the ejecta
(for SN IIP) plus an internal source (radioactive decays or magnetar spin-down) combined with the effects of H and He recombination. 
For SN IIP events where the H envelope is retained, the effects
of H recombination are more pronounced, while for SN Ib and SN Ic events the LCs lack an extended plateau phase due to the absence of
the outer envelope. Some SN Ib/c events develop SN IIn characteristics in their nebular spectra at late times,
indicating that the ejecta have finally reached the previously expelled H or He envelope and started to interact with it (Pastorello et al. 2008).

 On the other hand, in SN IIL and SN IIn events, the CSM envelopes have higher optical depth so that the effects of ejecta-CSM interaction
dominate the output luminosity. In the case of SN IIL the optical depth is probably moderate since P Cygni features are seen and
the LC decline is faster, while for SN IIn a high optical depth is implied that keeps the expansion of the SN ejecta obscured and produces
a LC of longer duration. The 
pronounced effects of the optically-thick CSM environment on the LC of luminous SN IIn manifest themselves by showing a variety
in the LC shapes and durations of these events. This leads us to argue that the SN IIL classification could be a byproduct of the
diversity of CSM properties and not a separate class of SN explosions. The faster decline and apparently symmetric shape of the LCs of some 
SN IIL and SLSNe (SCP06F6; Barbary et al. 2009, PTF09cwl, PTF09cnd and PTF09atu; Quimby et al. 2011)
may be well reproduced by models of shock heating, where the forward and reverse shocks terminate their contribution once they have swept up
most of the available SN ejecta and CSM mass.
If that is true, then strong pre-SN mass loss might constitute a significant source of
diversity among SNe IIn, producing SLSNe at the high mass end, and putting such
interacting SNe into a different parameter space compared to SNe that do not strongly interact.
This point of view was advocated by Blinnikov \& Bartunov (1993) where they presented radiation-hydrodynamics models of SN IIL
in order to fit the LC of SN~1979C.

We applied our hybrid CSM interaction plus radioactive decay model in the case of the LC of SLSN~2006gy. We found that the LC of this extraordinary
event could be reasonably well reproduced by a model where interaction of $\sim$~40~$M_{\odot}$ of SN ejecta with a circumstellar shell 
of $\sim$~5~$M_{\odot}$  plus radioactive
decay of $\sim$~2~$M_{\odot}$ of $^{56}$Ni provide the luminosity input. Although we derive a smaller $^{56}$Ni mass than previous authors, it remains the case
that SLSN~2006gy was a brilliant explosion and that the progenitor star must have undergone 
episodic mass-loss, a picture consistent with an LBV-type star (Smith et al. 2007). 
We note that alternative scenarios for the nature of the massive CSM shell around SLSN~2006gy, such
as a wind blown-bubble (Dwarkadas 2011) or a PPISN (Woosley, Heger \& Blinnikov; 2007) cannot be excluded.

The models we present here are only approximate and have a large number of parameters, but they also provide the means to efficiently
explore a large range of parameter space. In this way, parameters can be chosen to guide more elaborate and expensive numerical radiation-hydrodynamics
calculations.
In a subsequent paper we will present fits to the LCs
of many of the recently discovered SLSNe and other peculiar transient events. 
The ultimate goal is to deduce the physical properties of the progenitors and the CSM environments of a variety
of SNe using their observed optical LCs and spectra in order to constrain the mass-loss history and pre-SN evolution of massive stars.

We would like to thank Roger Chevalier, Vikram Dwarkadas, Lars Bildsten, Rob Robinson, Dave Arnett, Sean Couch,
and the anonymous referee for useful discussions and comments. We would especially like to thank Sergei Blinnikov and Takashi
Moriya for providing their results from radiation hydrodynamics simulations of SN ejecta CSM interaction that we have used
for benchmarking against our analytical models.
This research is supported in part by NSF Grants AST-0707669 and AST-1109801 and by the Texas Advanced Research 
Program grant ASTRO-ARP-0094. 
E. Chatzopoulos would like to thank the Propondis foundation of Piraeus, Greece for its support 
of his studies. J. Vinko received support from Hungarian OTKA Grant K76816.

%%%%%%%%%%%%%%%%%%%%%%%%%%%%%%%%%%%%%%%%%%%%%%%%%%%%%%%%%%%%%%%%%%%%%%%%%%%%
%%            REFERENCES
%%%%%%%%%%%%%%%%%%%%%%%%%%%%%%%%%%%%%%%%%%%%%%%%%%%%%%%%%%%%%%%%%%%%%%%%%%%%

{}                     
%\end{references}

\appendix

\section{Derivation of the general LC model for centrally located power sources.}

A80, A96 presented a solution for light curves powered by radioactive decay of $^{56}$Ni
based on the energy equation:
\begin{equation}
\dot{E}+P \dot{V}=\epsilon_{Ni}-\frac{\partial L}{\partial m},
\end{equation}
where $E$ is the specific internal energy, $P$ is the pressure, $V=\rho^{-1}$ is the specific
volume, $\epsilon_{Ni}$ is the specific energy generation rate corresponding to the
radioactive decay of $^{56}$Ni and the luminosity is given by the diffusion approximation:
\begin{equation}
L=-\frac{4 \pi r^{2} \lambda c a}{3} \frac{\partial T^{4}(x,t)}{\partial r},
\end{equation}
where $\kappa$ is an appropriate opacity. A80, 82 solved this set of equations
by assuming 1) homologous expansion, 2) a centrally-located power source, and
3) that radiation pressure was dominant. A80, 82 did not consider a general power
input source as we need to consider here.

Kasen \& Bildsten (2010) presented a related solution for the light curve based on an integrated version of the energy equation
and a general power input, of which a magnetar was a specific example. Kasen \& Bildsten did not show explicitly how
this global, integrated solution is related to the solution of A80, 82 that was expressed in terms of local, specific quantities. Here we
present the most general solution in the formulation of A80, 82, show how that solution is generalized to any power input, and how that
solution relates to the global, integrated solution. 

Following A80, 82 we adopt the same prescription for separation of variables,
\begin{equation}
T^{4}(x,t)=T_{00}^{4} \psi(x) \phi(t) [R_{0}/(R_{0}+v t)]^{4},
\end{equation}
where $x=r/R$ is the dimensionless radial variable, $T_{00}$ is the initial central temperature, $v$ is the
scaling velocity of the homologous expansion $v(x)=x v$, and $R_{0}$ is the initial radius.
Adopting $\kappa =$~constant, we can then write, following the analysis of A80, 82, the luminosity
at the surface, $x =$~1, as a function of time as:
\begin{equation}
L(1,t)=\frac{M \epsilon_{0} b I_{th}}{I_{M}} e^{-[\frac{t^{2}}{t_{d}^{2}}+\frac{2R_{0}t}{vt_{d}^{2}}]}
\int_0^t e^{[\frac{t'^{2}}{t_{d}^{2}}+\frac{2R_{0}t'}{vt_{d}^{2}}]} f(t')
[\frac{R(t')}{t_{0}R_{0}}]dt' + \frac{E_{th,0} I_{M}}{M \epsilon_{0} b I_{th}} e^{-(t^{2}/t_{d}^{2}+2R_{0}t/vt_{d}^{2})},
\end{equation}
where $M$ is the ejecta mass, $\epsilon_{0}$ is the amplitude of the specific power input in erg~$g^{-1}$~$s^{-1}$, $f(t)$ is
the time-dependence of the power input, $E_{th,0} = 4 \pi R_{0}^{3} a T_{00}^{4} \phi(0) I_{th}$ is the initial total thermal energy,
$t_{d}$ is the LC time-scale which is the geometric mean of the diffusion time-scale $t_{0} \equiv 3 \kappa R_{0}^{2} \rho_{00}/c \alpha$ 
and the expansion time-scale $t_{h} = R_{0}/v$, therefore $t_{d}^{2}=2 t_{0} t_{h}$.
The factor $R(t)/R_{0}t_{0}$ can be
written as $(R_{0}+vt)/R_{0}t_{0}$ or $(1/t_{d})(2R_{0}/vt_{d}+2t/t_{d})$, as convenient to evaluate the limits $R_{0} \rightarrow$~0
and $v \rightarrow$~0.
The density is assumed to scale as $\rho(x,t)=\rho_{00}\eta(x)[R_{0}/R(t)]^{3}$, where $\rho_{00}$ is the central density at time 0 and
the function $\alpha$ appearing in the definition of $t_{0}$,
\begin{equation}
\alpha \equiv -\frac{1}{x^{2}\psi(x)}\frac{\partial}{\partial x}[\frac{x^{2}}{\eta(x)}\frac{\partial \psi}{\partial x}],
\end{equation}
is a constant by separation of variables. The parameter $b = \xi(x)\eta(x)/\psi(x)$ is assumed to be constant where $\xi(x)$ represents
the radial distribution of the power input, $I_{th} \equiv \int_{0}^{1} \psi(x)x^{2}dx$, and $I_{M} \equiv \int_{0}^{1} \eta(x)x^{2}dx$. A80, 82
show that $t_{0} = \kappa M/\beta c R_{0}$ where $\beta \equiv 4 \pi \alpha I_{M}/3 \simeq 13.8$ for a variety of density distributions and
that $bMI_{th}/I_{M}=M_{Ni}^{0}$, the initial amount of nickel injected in the radioactive decay model.

With these definitions and relations we can write for a general power input
\begin{equation}
\epsilon_{inp} \equiv \epsilon_{0} \xi(x) f(t),
\end{equation}
and, expressed in terms of luminosity 
\begin{equation}
L_{inp}=\int_{0}^{M} \epsilon_{inp} dm = L_{0} f(t),
\end{equation}
where
\begin{equation}
L_{0} = M \epsilon_{0} b \frac{I_{th}}{I_{M}},
\end{equation}
or, formally,
\begin{equation}
L(1,t)=L_{0} e^{-[\frac{t^{2}}{t_{d}^{2}}+\frac{2R_{0}t}{vt_{d}^{2}}]}
\int_0^t e^{[\frac{t'^{2}}{t_{d}^{2}}+\frac{2R_{0}t'}{vt_{d}^{2}}]} f(t')
[\frac{R(t')}{t_{0}R_{0}}]dt' + \frac{E_{th,0}}{t_{0}} e^{-(t^{2}/t_{d}^{2}+2R_{0}t/vt_{d}^{2})}.
\end{equation}
This is the form of the output luminosity we present in Equation 3 of the main text.

Returning to Equation A1, we can integrate the energy equation over mass to find
\begin{equation}
4 \pi R(t)^{3} a T_{00}^{4} [\frac{R_{0}}{R(t)}]^{4} \dot{\phi}  I_{th} = L_{inp} - L
\end{equation}
where $L=\int_{0}^{M} (\partial L/\partial m)dm$.
Kasen \& Bildsten (2010) began with an approximate integrated form of the energy equation to write
\begin{equation}
\frac{d}{dt} [\frac{4}{3} \pi R(t)^{3} a T^{4}] + \frac{1}{3} a T^{4} 4 \pi R^{2} \frac{dR(t)}{dt} = L_{inp}-\frac{4 \pi R(t)^{2} c}{3 \kappa
\rho}\frac{\partial a T^{4}}{\partial r},
\end{equation}
where the second term on the right hand side is $- L$.
Using the same separation of variables for $T$ as in Equation A3, this equation can be written as
\begin{equation}
4 \pi R(t)^{3} a T_{00}^{4} [\frac{R_{0}}{R(t)}]^{4} \dot{\phi} \frac{\psi(x)}{3} = L_{inp} - L.
\end{equation}
This equation agrees with Equation A10 only if $\psi(x) = 3I_{th}$, which is only true if $\psi =$~constant. This means
that the temperature as a function of radius has to be constant. The solution of Kasen \& Bildsten is thus not strictly speaking
self-consistent, since a temperature gradient is required and employed to estimate the luminosity. Their solution is nevertheless a useful approach, given
the uncertainties involved in modeling light curves driven by uncertain phenomena.

\section{Derivation of the SN ejecta-CSM interaction self-similar luminosity input.}

To derive an expression for the luminosities of the forward and reverse shocks we assume that all of their kinetic energy
converts efficiently to radiation (radiative shock approximation):
\begin{equation}
L=\frac{dE}{dt}=\frac{d}{dt}\left(\frac{1}{2}M_{sw}v_{sh}^{2}\right)=M_{sw} v_{sh} \dot{v_{sh}}
+\frac{1}{2} \dot{M_{sw}} v_{sh}^{2},
\end{equation}
where $M_{sw}$ is the swept-up mass behind the shock and $v_{sh}$ the shock velocity.

In order to estimate a final expression for the luminosity presented in Equation B1, we need to know
the dynamics of the forward and reverse shock (radius, velocity and acceleration) as a function of time. Chevalier (1982) considered 
the interaction between two media with power-law density profiles: the SN ejecta density profile
$\rho_{SN}=g^{n} t^{n-3}r^{-n}$, where $g^{n}$ is a scaling parameter for the ejecta density profile, 
$g^{n} = 1/(4 \pi (\delta-n)) [2(5-\delta)(n-5)E_{SN}]^{(n-3)/2}/[(3-\delta)(n-3)M_{ej}]^{(n-5)/2}$, $n$ is the power-law exponent and 
$\delta$ is the slope of the inner density profile of the ejecta (values of $\delta =$~0, 2 are typical) and the CSM density profile $\rho_{CSM}=q r^{-s}$, 
where $\rho_{CSM}$ is the density of the CSM medium, $q$ is a scaling constant and $s$ the power-law exponent.
In general $q=\rho_{CSM,1} r_{1}^{s}$ where $\rho_{CSM,1}$ is the density of the CSM shell at $r=r_{1}$. We use as fiducial value
$r_{1} = R_{p}$, where $R_{p}$ the radius of the progenitor star. Thus we set the density scale of the CSM, $\rho_{CSM,1}$, immediately
outside the stellar envelope.

Using momentum conservation, Chevalier (1982) and Chevalier \& Fransson (2001) 
found the following self-similar solutions for the radii of the forward and the reverse shocks, respectively, as a function of time:
\begin{equation}
R_{F}(t)=R_{p}+\beta_{F} \left[\frac{A g^{n}}{q}\right]^{\frac{1}{n-s}}t^{\frac{n-3}{n-s}},
\end{equation}
and
\begin{equation}
R_{R}(t)=R_{p}+\beta_{R} \left[\frac{A g^{n}}{q}\right]^{\frac{1}{n-s}}t^{\frac{n-3}{n-s}},
\end{equation}
where $\beta_{F}$, $\beta_{R}$ and $A$ are constants that depend on the values of $n$ and $s$ and, for a variety of values, are
given in Table 1 of Chevalier (1982). More specifically the parameters $\beta_{F}$ and $\beta_{R}$ refer to the ratio of the shock radius
to the radius of the contact discontinuity that forms as a result of SN ejecta-CSM interaction ($\beta_{F}=R_{1}/R_{c}$ and 
$\beta_{R}=R_{2}/R_{c}$ for the forward and the reverse shock, respectively, where $R_{c}$ is the radius of the contact discontinuity given by 
Equation 3 of Chevalier (1982) and $R_{1}$ and $R_{2}$ are the radii of the forward and the reverse shocks respectively).
Using $v_{F,R}=dR_{F,R}/dt$, $\dot{v_{F,R}}=d^{2}R_{F,R}/dt^{2}$ yields the velocity and the acceleration of the shock
as a function of time.

Given the shock dynamics derived above, we can calculate the swept-up mass behind the forward shock assuming that $R_{p}$
is smaller than $R_{F}(t)$ for $t>0$:
\begin{equation}
M_{sw,F}(t)=4\pi \int_{R_{p}}^{R_{F}(t)} \rho_{CSM}(r)r^{2}dr = 
\frac{4 \pi \beta_{F}^{3-s}}{3-s} q^{\frac{n-3}{n-s}}[A g^{n}]^{\frac{3-s}{n-s}} t^{\frac{(n-3)(3-s)}{n-s}},
\end{equation}
and the swept-up mass behind the reverse shock:
\begin{equation}
M_{sw,R}(t)=4\pi \int_{R_{R}(t)}^{R_{SN}(t)} \rho_{SN}(r)r^{2}dr = 
\frac{4 \pi g^{n} v_{SN}^{3-n}}{3-n} 
(1-\frac{\beta_{R} \left[\frac{A g^{n}}{q}\right]^{\frac{1}{n-s}}t^{\frac{n-3}{n-s}}}{v_{SN} t})^{3-n},
\end{equation}
where $v_{SN}$ is the characteristic SN expansion velocity and $R_{SN}=v_{SN} t$ is the radius of the SN photosphere which is assumed
to expand homologously. We have again assumed that $R_{p}$ is small.

Substituting Equations B2 through B5 and their derivatives in Equation B1 yields the final results for the luminosity input
from the forward and the reverse shocks ($L_{F}(t)$ and $L_{R}(t)$ respectively):
\begin{equation}
L_{F}(t)=\frac{2 \pi}{(n-s)^{3}} g^{n\frac{5-s}{n-s}} q^{\frac{n-5}{n-s}} (n-3)^{2} (n-5) \beta_{F}^{5-s} A^{\frac{5-s}{n-s}}
(t+t_{i})^{\frac{2n+6s-ns-15}{n-s}} \theta(t_{FS,BO}-t),
\end{equation}
and
\begin{equation}
L_{R}(t)=2 \pi (\frac{A g^{n}}{q})^{\frac{5-n}{n-s}} g^{n} (\frac{3-s}{n-s})^{3} 
(t+t_{i})^{\frac{2n+6s-ns-15}{n-s}} \theta(t_{RS,*}-t),
\end{equation}
where $\theta(t_{FS,*}-t)$ and $\theta(t_{RS,*}-t)$ denote the heaviside step function that
controls the termination of the forward and reverse shock, respectively, and 
$t_{i} \simeq R_{p}/v_{SN}$ is the initial time of the
CSM interaction which sets the initial value for the luminosity produced. The velocity, $v_{SN}$, in Equation B5 is given by
\begin{equation}
v_{SN} = \frac{[10(n-5)E_{SN}/3(n-3)M_{ej}]^{\frac{1}{2}}}{x_{0}},
\end{equation}
where $x_{0}=r_{0}(t)/R_{SN}(t)$ is the dimensionless radius
of the break in the SN ejecta density profile from the inner flat component (controlled by $\delta$) to the outer, steeper component
(controlled by $n$) which is at radius $r_{0}(t)$. $L_{R}(t)$ in Equation B7 is independent of $v_{SN}$ and hence of $x_{0}$.
Summing Equations B6 and B7, the total luminosity input from self-similar SN ejecta-CSM interaction is
\begin{equation}
L_{inp}(t)=L_{F}(t)+L_{R}(t).
\end{equation}
The forward shock termination time-scale, $t_{FS,*}$, is given by the following expression, assuming
that the input from the forward shock terminates when all the available CSM has been swept up:
\begin{equation}
t_{FS,*}=|\frac{(3-s)q^{(3-n)/n-s)} [Ag^{n}]^{(s-3)/(n-s)}}{4 \pi \beta_{F}^{3-s}}|^{\frac{n-s}{(n-3)(3-s)}} M_{CSM}^{\frac{n-s}{(n-3)(3-s)}},
\end{equation}
where $M_{CSM}$ the total mass of the CSM. Following the same assumption, the reverse shock termination time-scale $t_{RS,*}$ is given by the 
following expression:
\begin{equation}
t_{RS,*}=[\frac{v_{SN}}{\beta_{R} (Ag^{n}/q)^{\frac{1}{n-s}}} (1-\frac{(3-n)M_{ej}}{4\pi v_{SN}^{3-n} g^{n}})^{\frac{1}{3-n}}]^{\frac{n-s}{s-3}}.
\end{equation}
After termination, the luminosity of each component decays in a manner analogous to Equation 6 in the main text such that
\begin{equation}
L_{F}(t)=L_{F}(t_{FS,BO}) e^{-(t-t_{FS,BO})/t_{0}},
\end{equation}
and
\begin{equation}
L_{R}(t)=L_{R}(t_{RS,*}) e^{-(t-t_{RS,*})/t_{0}}.
\end{equation}

%%%%%%%%%%%%%%%%%%%%%%%%%%%%%%%%%%%%%%%%%%%%%%%%%%%%%%%%%%%%%%%%%%%%%%%%%%%%
%%            FIGURES 

\begin{figure}
\begin{center}
\includegraphics[angle=-90,width=13cm]{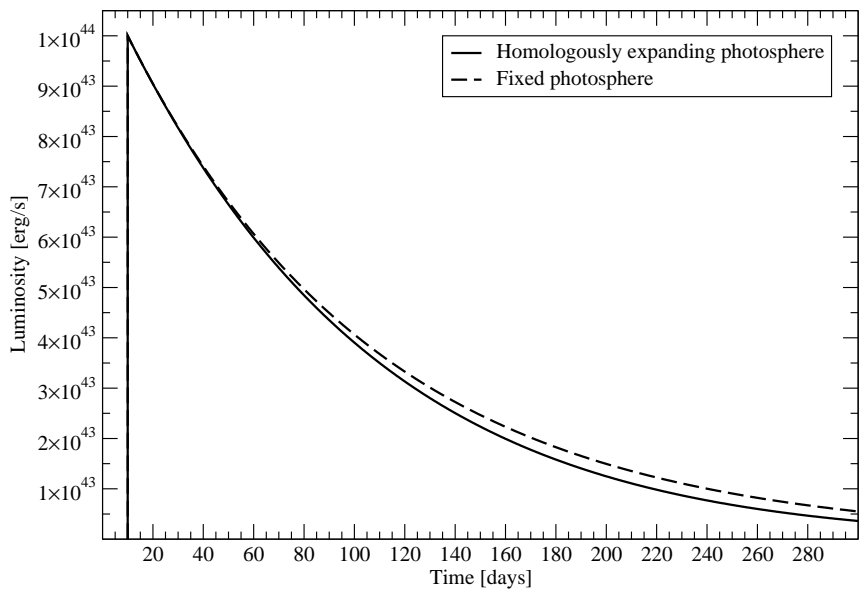}
\caption{Examples of LC models for instantaneous luminosity input in the cases of
homologously expanding (solid curve) and fixed (dashed curve) photosphere (see \S 3.1).
The decline is determined purely by diffusion. 
The initial luminosity
for this model is $L_{0} =$~$10^{44}$~erg~s$^{-1}$ introduced at $t =$~10~days for the choice of diffusion
time $t_{0} =$~100~days. 
For the homologously expanding photosphere case the initial radius is $R_{0} = 10^{12}$~cm and the photospheric
velocity $v =$~10,000~km~s$^{-1}$. For the fixed photosphere, $R_{0} =$~10$^{12}$~cm=constant.}
\end{center}
\end{figure}

\begin{figure}
\begin{center}
\includegraphics[angle=-90,width=13cm]{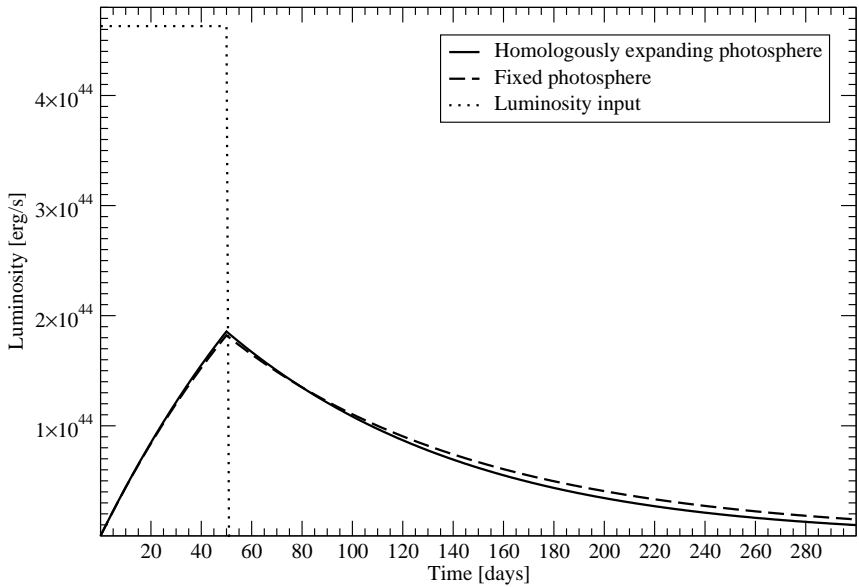}
\caption{Example of LC models for a terminated constant luminosity input (``top-hat" - dotted curve) 
in the cases of homologously expanding matter (solid curve) and fixed photosphere (dashed curve). See \S 3.2.1 for details. 
The models shown are for total input energy $2 \times 10^{51}$~erg that terminates at $t_{sh} =$~50~days and
for the choice of diffusion time $t_{0} =$~100~days. 
For the homologously expanding case the initial radius is $R_{0} = 10^{12}$~cm and the photospheric
velocity $v =$~10,000~km~s$^{-1}$. 
For the fixed photosphere, $R_{0} =$~10$^{12}$~cm=constant.
For this choice of parameters the diffusion mass is determined to be 11.6~$M_{\odot}$.}
\end{center}
\end{figure}

\begin{figure}
\begin{center}
\includegraphics[angle=-90,width=13cm]{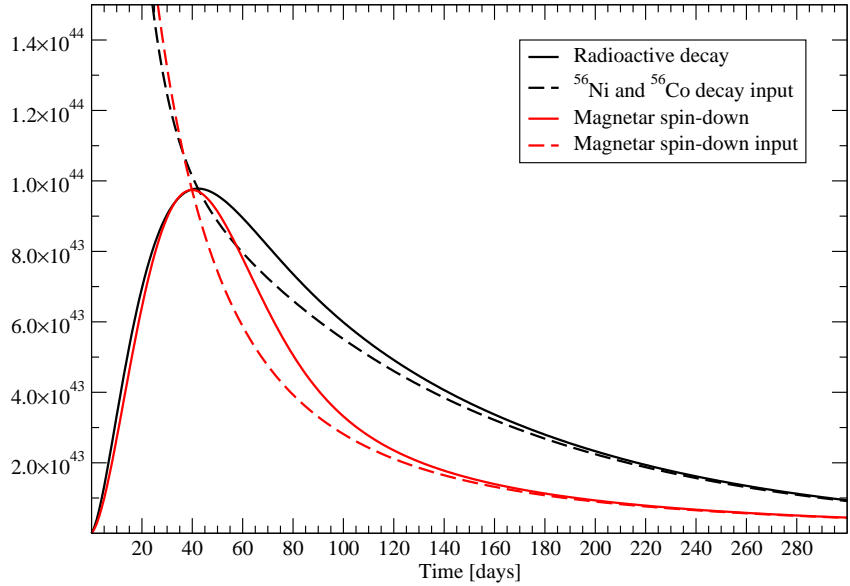}
\caption{Example of LC models for $^{56}$Ni and $^{56}$Co radioactive decay input (dashed black curve) 
and for magnetar spin-down input (dashed red curve; the output LC models are presented with solid curves
of the same corresponding colors).
The radioactive decay diffusion model is drawn for $M_{Ni} =$~10~$M_{\odot}$ and the magnetar spin-down model
for $E_{p} =$~$1.37 \times 10^{51}$~erg and $t_{p}$ =~30~days. The choices of the model
parameters were made such that the two model LCs have approximately the same maximum luminosity.
Both models are drawn for the homologously expanding photosphere case for diffusion time $t_{0} =$~40~days, 
$R_{0} = 10^{14}$~cm and $v =$~10,000~km~s$^{-1}$. For details on the models see \S 3.2.}
\end{center}
\end{figure}

\begin{figure}
\begin{center}
\includegraphics[angle=-90,width=13cm]{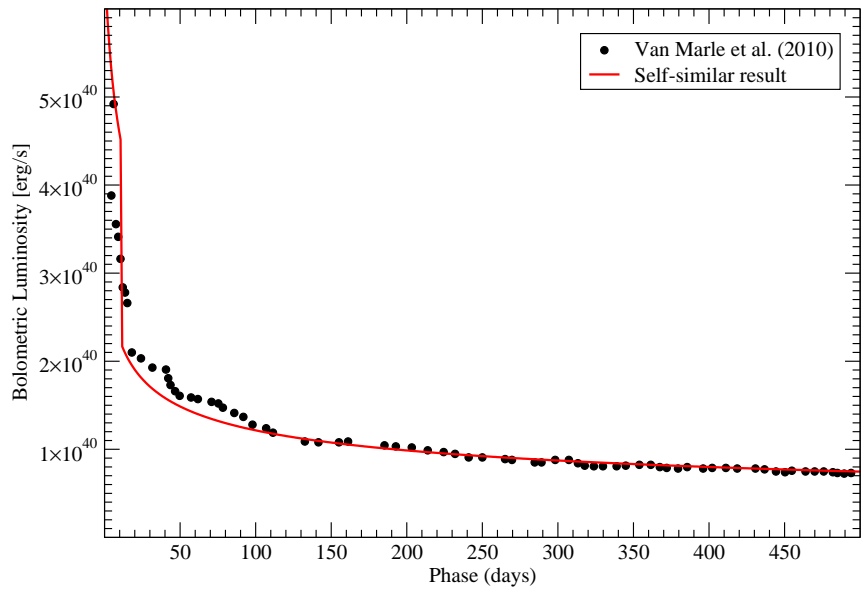}
\caption{Comparison of the model LC from the Van Marle et al. (2010) two-dimensional optically-thin radiation 
hydrodynamics simulation of the collision of SN ejecta from a red supergiant progenitor with a steady-state wind ($s=2$) CSM component 
(filled circles) with the analytic self-similar result for the FS luminosity for the same CSM and SN ejecta parameters. The break at about 10 days
in the self-similar model represents the time of the forward shock termination.}
\end{center}
\end{figure}

\begin{figure}
\begin{center}
\includegraphics[angle=-90,width=13cm]{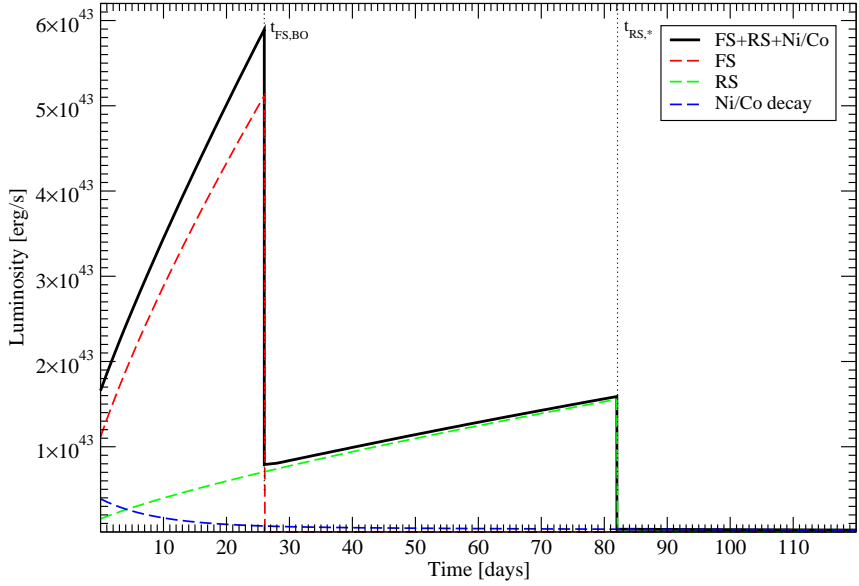}
\caption{Example of a hybrid ejecta-CSM interaction plus $^{56}$Ni and $^{56}$Co radioactive decay input (solid
black curve). The dashed red, green and blue curves represent the forward, reverse and radioactive decay luminosity inputs, respectively.
The vertical dotted lines indicate the termination time for the forward and reverse shock luminosity inputs.
This hybrid luminosity input model is drawn for the following choice of parameters: $\delta =$~0, $n =$~12, $s =$~0, $E_{SN} = 
1 \times 10^{51}$~erg, $M_{ej} =$~15~$M_{\odot}$, $M_{CSM} =$~1~$M_{\odot}$, $R_{p} =$~$1 \times 10^{14}$~cm,
$\rho_{CSM,1} =$~$5 \times 10^{-13}$~g~cm$^{-3}$, $M_{Ni} =$~0.05~$M_{\odot}$.}
\end{center}
\end{figure}

\begin{figure}
\begin{center}
\includegraphics[angle=-90,width=13cm]{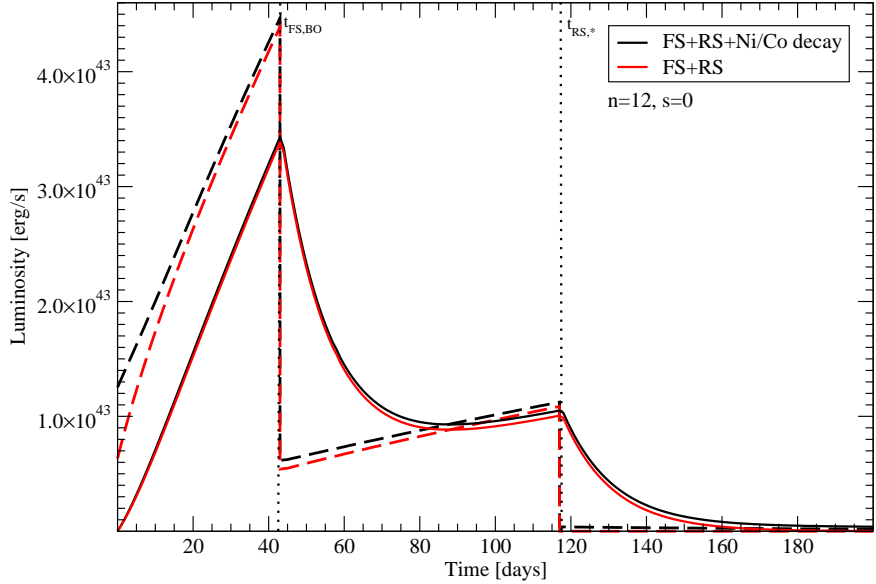}
\caption{Comparison of SN ejecta-CSM interaction and hybrid SN ejecta-CSM interaction plus $^{56}$Ni and 
$^{56}$Co radioactive decay LC models for the case of collision with a constant density CSM shell ($n =$~12, $s =$~0).
The hybrid forward, reverse and $^{56}$Ni and $^{56}$Co radioactive decay model is shown in black and the pure
forward and reverse shock ejecta-CSM interaction model in red.
In each case the dashed curve represents the input luminosity and the solid curve the output LC model.
The dotted vertical lines indicate the forward shock breakout time (essentially equal
to the time of termination of forward shock input in the optically-thick part of the CSM)
and the reverse shock termination time scale.
The forward termination time scale is very close to its breakout time scale in the case of a CSM constant density shell. 
The models were drawn for the following choice of parameters: $\delta =$~0, 
$E_{SN} =  10^{51}$~erg, $M_{ej} =$~12~$M_{\odot}$, $M_{CSM} =$~1~$M_{\odot}$, $R_{p} =$~$10^{14}$~cm,
$\rho_{CSM,1} =$~$10^{-13}$~g~cm$^{-3}$, $M_{Ni} =$~0.08~$M_{\odot}$.}
\end{center}
\end{figure}

\begin{figure}
\begin{center}
\includegraphics[angle=-90,width=13cm]{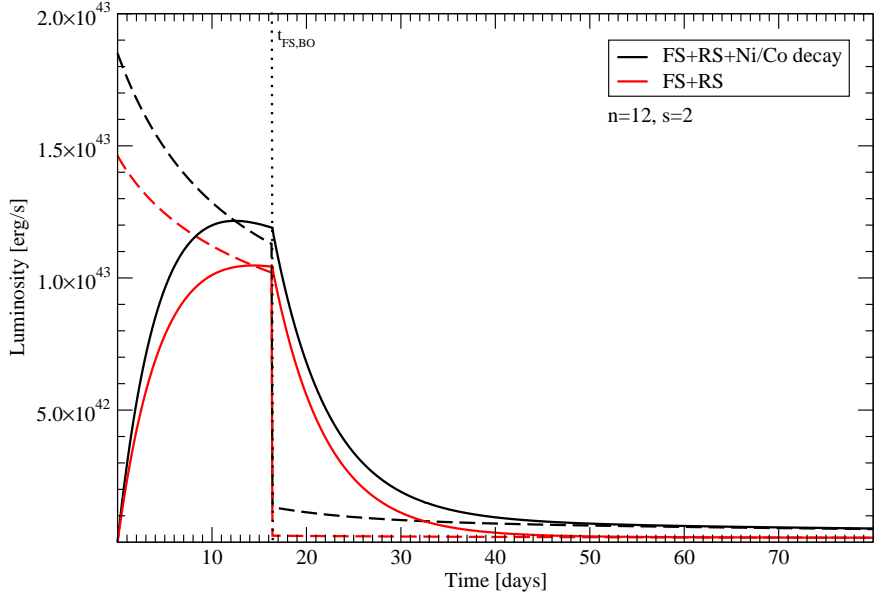}
\caption{Comparison of SN ejecta-CSM interaction and hybrid SN ejecta-CSM interaction plus $^{56}$Ni and 
$^{56}$Co radioactive decay LC models for the case of collision with a steady state wind ($n =$~12, $s =$~2).
The wind is terminated at a finite mass.
The hybrid forward, reverse and $^{56}$Ni and $^{56}$Co radioactive decay model is shown in black and the pure
forward and reverse shock ejecta-CSM interaction model in red.
In each case the dashed curve represents the input luminosity and the solid curve the output LC model.
The dotted vertical lines indicate the forward shock breakout time (essentially equal 
to the time of termination of forward shock input in the optically-thick part of the CSM)
and the reverse shock termination time scale.
The reverse shock termination time scale is much later, outside of the range of this graph.
The models were drawn for the following choice of parameters: $\delta =$~0, 
$E_{SN} = 10^{51}$~erg, $M_{ej} =$~12~$M_{\odot}$, $M_{CSM} =$~0.1~$M_{\odot}$, $R_{p} =$~$2 \times 10^{14}$~cm,
$\rho_{CSM,1} =$~$5 \times 10^{-13}$~g~cm$^{-3}$, $M_{Ni} =$~0.05~$M_{\odot}$.}
\end{center}
\end{figure}

\begin{figure}
\begin{center}
\includegraphics[angle=-90,width=13cm]{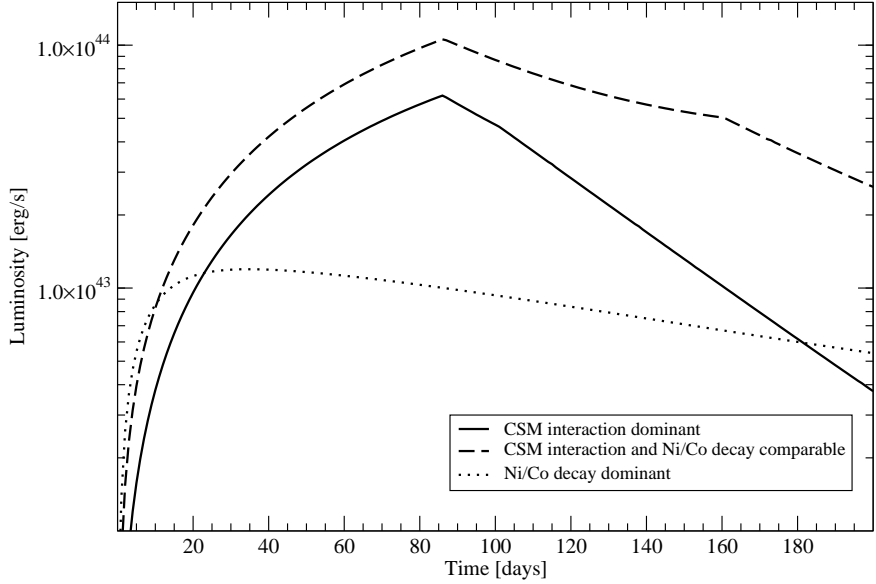}
\caption{Comparison of hybrid CSM interaction plus radioactive decay LC models for a case where CSM interaction in a constant density shell is the dominant
contributor to the output luminosity (solid curve), a case where CSM interaction and radioactive decay have comparable luminosity inputs
(dashed curve) and a case where radioactive decay is the dominant input luminosity source (dotted curve). The post-maximum decline rate
can be more rapid for the hybrid models for which CSM interaction dominates or is comparable to the radioactive decay input. Such models are consistent
with the decay seen in SN IIL. For appropriate choices of parameters, the hybrid model can also produce an approximately symmetric LC
shape around peak luminosity, similar to the observed optical LC of the peculiar transient SCP~06F6 (Barbary et al. 2008) as illustrated here by the model
for which the CSM dominates.
For the case where CSM interaction is dominant the following parameters were used:
$\delta =$~0, $n =$~12, $s =$~0,
$E_{SN} = 10^{51}$~erg, $M_{ej} =$~10~$M_{\odot}$, $M_{CSM} =$~5~$M_{\odot}$, $R_{p} =$~$110^{14}$~cm,
$\rho_{CSM,1} =$~$10^{-13}$~g~cm$^{-3}$, $M_{Ni} =$~0.05~$M_{\odot}$.
For the case where radioactive decay input is comparable to the CSM input the following parameters were used:
$\delta =$~0, $n =$~12, $s =$~0,
$E_{SN} = 4 \times 10^{51}$~erg, $M_{ej} =$~40~$M_{\odot}$, $M_{CSM} =$~7~$M_{\odot}$, $R_{p} =$~$10^{14}$~cm,
$\rho_{CSM,1} =$~$10^{-13}$~g~cm$^{-3}$, $M_{Ni} =$~2~$M_{\odot}$.
Finally, for the case where the radioactive decay input is dominant the following parameters were used:
$\delta =$~0, $n =$~12, $s =$~0,
$E_{SN} = 7 \times 10^{50}$~erg, $M_{ej} =$~21~$M_{\odot}$, $M_{CSM} =$~2~$M_{\odot}$, $R_{p} =$~$10^{14}$~cm,
$\rho_{CSM,1} =$~$6 \times 10^{-16}$~g~cm$^{-3}$, $M_{Ni} =$~0.8~$M_{\odot}$.}
\end{center}
\end{figure}

\begin{figure}
\begin{center}
\includegraphics[angle=-90,width=13cm]{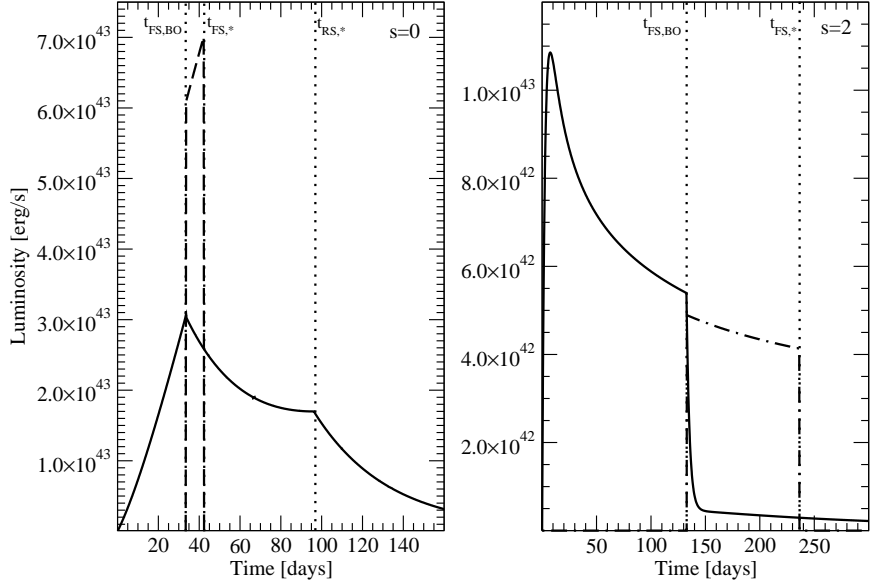}
\caption{Comparison between the optical (solid curve) and the UV/X-ray (dashed curve) LC for a hybrid CSM interaction
plus $^{56}$Ni and $^{56}$Co radioactive decay model in the case of SN ejecta interaction with a shell ($s =$~0; left panel) and a wind 
($s =$~2; right panel). 
The dotted vertical lines indicate the times of forward shock breakout (essentially equal
to the time of termination of forward shock input in the optically-thick part of the CSM),
forward shock termination (the end of optically-thin X-ray and UV input by the forward shock)
and reverse shock termination.
The $s =$~0 model corresponds to the following choice of parameters: $\delta =$~0, $n =$~12,
$E_{SN} = 1.2 \times 10^{51}$~erg, $M_{ej} =$~20~$M_{\odot}$, $M_{CSM} =$~2~$M_{\odot}$, $R_{p} =$~$2 \times 10^{14}$~cm,
$\rho_{CSM,1} =$~$5 \times 10^{-13}$~g~cm$^{-3}$, $M_{Ni} =$~0.05~$M_{\odot}$.
The $s =$~2 model corresponds to the following choice of parameters: $\delta =$~0, $n =$~12
$E_{SN} = 1.2 \times 10^{51}$~erg, $M_{ej} =$~18~$M_{\odot}$, $M_{CSM} =$~1~$M_{\odot}$, $R_{p} =$~$10^{14}$~cm,
$\rho_{CSM,1} =$~$2 \times 10^{-12}$~g~cm$^{-3}$, $M_{Ni} =$~0.05~$M_{\odot}$.}
\end{center}
\end{figure}

\begin{figure}
\begin{center}
\includegraphics[angle=-90,width=13cm]{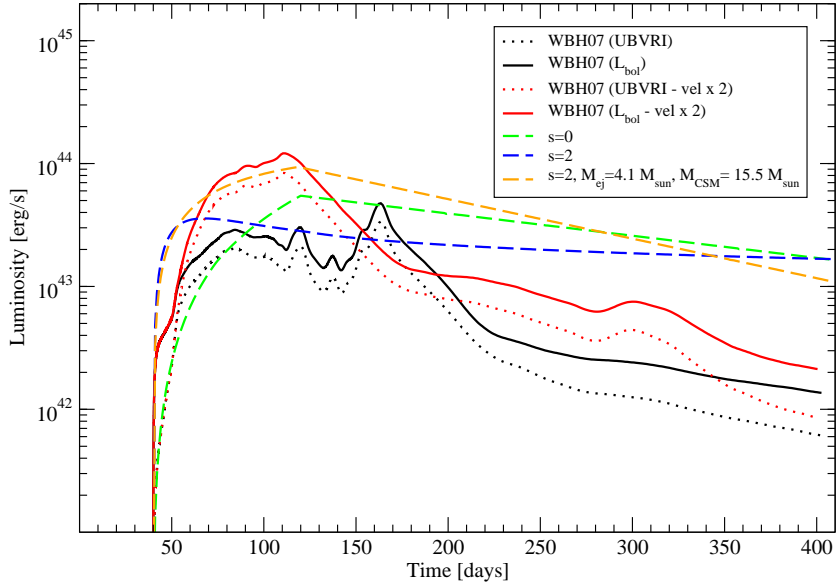}
\caption{Comparison of analytical SN LCs with numerical results from Woosley, Blinnikov \& Heger (2007) for the same
initial conditions. The solid black and red curves correspond to the bolometric LCs for the original ejecta velocity and a doubled
ejecta velocity, respectively, and the dotted black and red curves correspond to the UBVRI LCs as calculated by WBH07. The dashed curves show our
analytical LCs for $s =$~0 (green curve), for $s =$~2 (blue curve) and for $s =$~2 but for the values of $M_{ej}$ and $M_{CSM}$ 
adjusted in order to provide a better fit to the results from the simulations (yellow curve; see text).}
\end{center}
\end{figure}

\begin{figure}
\begin{center}
\includegraphics[angle=-90,width=13cm]{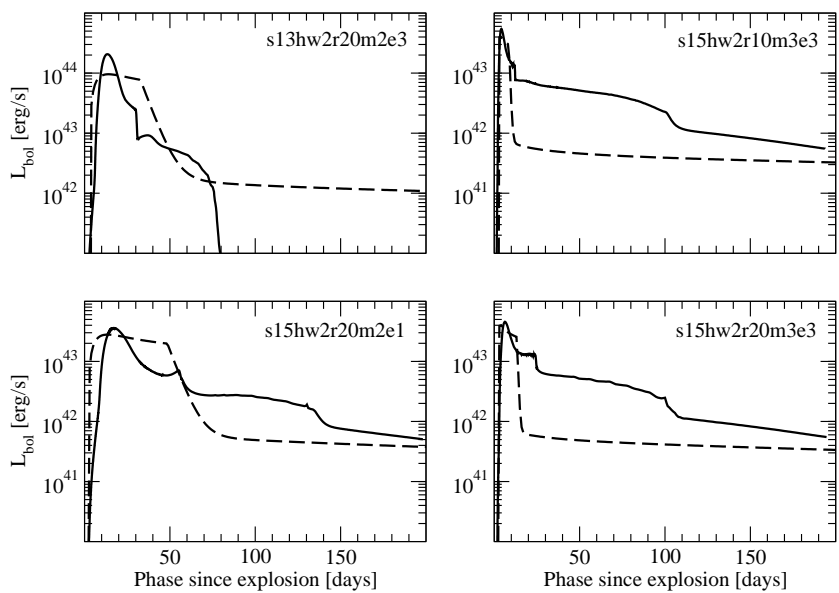}
\caption{Comparison of analytical SN LCs (dashed curves) with numerical results from Moriya et al. (2011) (solid curves) 
for the same set of initial conditions for four models: s13hw2r20m2e3 (upper left panel), s15hw2r10m3e3 (upper right panel), 
s15hw2r20m2e1 (lower left panel) and s15hw2r20m3e3 (lower right panel). See text for model parameters.}
\end{center}
\end{figure}

\begin{figure}
\begin{center}
\includegraphics[angle=-90,width=13cm]{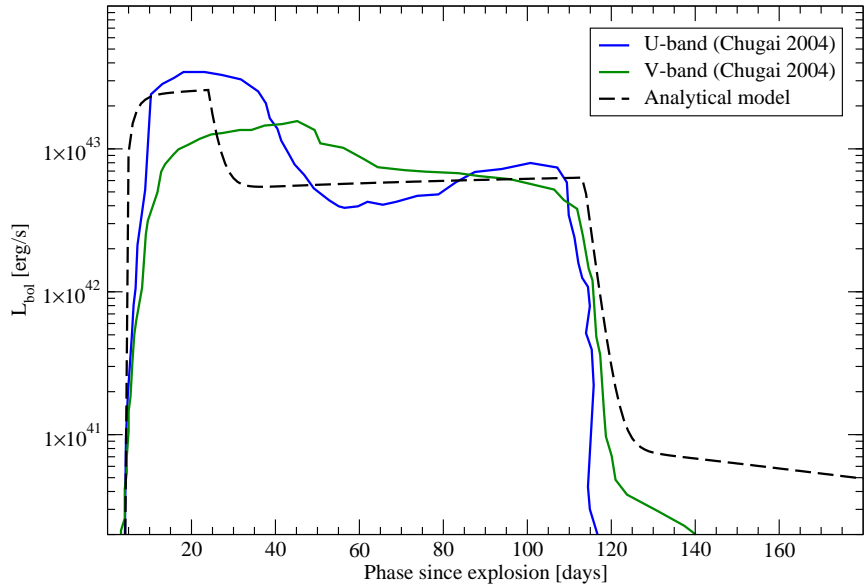}
\caption{Comparison of analytical SN LC (dashed curve) with the numerical
V-band (solid green curve) and U-band (solid blue curve) LCs for model sn94w58 of Chugai et al. (2004) for SN~1994W. See text for model parameters.}
\end{center}
\end{figure}

\begin{figure}
\begin{center}
\includegraphics[angle=-90,width=13cm]{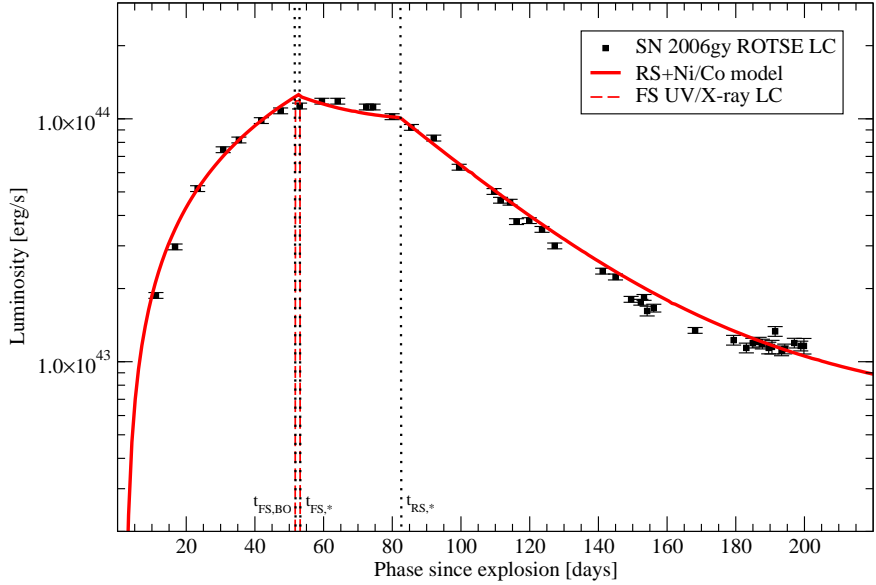}
\caption{Fit of a hybrid ejecta-CSM interaction plus $^{56}$Ni and $^{56}$Co radioactive decay LC model (solid red curve) to the ROTSE LC
of SLSN~2006gy. The parameters for this model are the following: $\delta =$~0, $n =$~12, $s =$~0,  
$E_{SN} = 4.4 \times 10^{51}$~erg, $M_{ej} =$~40~$M_{\odot}$, $M_{CSM} =$~5~$M_{\odot}$, $R_{p} =$~$5 \times 10^{14}$~cm,
$\rho_{CSM,1} =$~$1.5 \times 10^{-13}$~g~cm$^{-3}$, $M_{Ni} =$~2~$M_{\odot}$.
The dashed curve shows the expected UV/X-ray LC for this event corresponding to those fitting parameters. 
The dotted vertical lines indicate the times of forward shock breakout (essentially equal
to the time of termination of forward shock input in the optically-thick part of the CSM),
forward shock termination (the end of optically-thin X-ray and UV input by the forward shock) and reverse shock termination.}
\end{center}
\end{figure}

%%%%%%%%%%%%%%TABLE 1%%%%%%%%%%%%%%%%%%%%%%%%%%%%%%%%%%%%%%%%%%%%%%%%%%%%%%%

\end{document}